%% file: paper.tex
\documentclass[10pt,onecolumn,draftcls]{IEEEtran} 
\usepackage{verbatim} 
\usepackage{mathtools}
\usepackage{amsfonts,amsmath,mathrsfs} 
\usepackage[final]{graphicx} 
\usepackage{color} 
\usepackage{psfrag, times} 
\usepackage{amssymb,amsbsy,amsthm} 
\usepackage{subfigure} 
\usepackage{algorithmic,algorithm} 
\usepackage[latin1]{inputenc} 
\usepackage{tikz} 
\usepackage{epstopdf}
\usetikzlibrary{calc,trees,positioning,arrows,chains,shapes.geometric,%
    decorations.pathreplacing,decorations.pathmorphing,shapes,%
    matrix,shapes.symbols} 
 
\begin{document} 
\include{notation} 
\title{Degrees of Freedom of Interference Channels with CoMP Transmission and Reception} 
\author{{\large{V. Sreekanth Annapureddy, {\em Student Member, IEEE}, Aly El Gamal, {\em Student Member, IEEE}, and Venugopal V. Veeravalli, {\em Fellow, IEEE}}}
\thanks{The authors are with the Coordinated Science Laboratory and the
Department of Electrical and Computer Engineering,
University of Illinois at Urbana-Champaign, Urbana, IL 61801 USA (e-mail: sreekanthav@gmail.com, alyelgamal@gmail.com,vvv@illinois.edu).}
\thanks{This paper was presented in part at the International Symposium on Information Theory (ISIT), Austin, TX, July 2010.}
\thanks{This research was supported in part by the NSF award CCF-0904619, through the University of Illinois at Urbana-Champaign, and grants from Intel and Motorola Solutions. }}
\maketitle 
\begin{abstract} 
We study the Degrees of Freedom (DoF) of the $K$-user interference channel with coordinated multi-point (CoMP) transmission and reception. Each message is jointly transmitted by $M_t$ successive transmitters, and is jointly received by $M_r$ successive receivers. We refer to this channel as the CoMP channel with a transmit cooperation order of $M_t$ and receive cooperation order of $M_r$. Since the channel has a total of $K$ transmit antennas and $K$ receive antennas, the maximum possible DoF is equal to $K$. We show that the CoMP channel has $K$ DoF if and only if  $M_t + M_r \geq K+1$. For the general case, we derive an outer bound that states that the DoF is bounded above by $\left\lceil(K+M_t+M_r-2)/2\right\rceil$. For the special case with only CoMP transmission, i.e, $M_r = 1$, we propose a scheme that can achieve $(K+M_t-1)/2$ DoF for all $K < 10$, and conjecture that the result holds true for all $K$ . The achievability proofs are based on the notion of algebraic independence from algebraic geometry. 
\end{abstract} 
\section{Introduction} 
Interference is identified as a major bottleneck in realizing a ubiquitous and high-speed wireless world. There has been considerable interest in understanding the best ways to manage interference in wireless networks. Recent progress \cite{cadambe2008interference,OneBit2008,bresler2008two,Shang-Kramer-Chen-IT-2008,Motahari-Khandani-IT-2008,Annapureddy-Veeravalli-IT-2008} on Gaussian interference channels has advanced our understanding of the fundamental limits of communication in the presence of interference. The Gaussian interference channel has a finite (say $K$) number of transmitter-receiver pairs with each transmitter having a message desired by the respective receiver. Among other settings, the interference channel is a good model for cellular wireless networks, both downlink and uplink.  Even if we can determine and implement the best possible achievable schemes for the interference channel, the demand for wireless connectivity is likely to exceed what the physical channel can offer.  For this and other reasons, there has been much interest in understanding the fundamentals limits of cooperative interference networks. Typically cooperation requires additional infrastructure, but it could be cost-effective depending on the overall objective.  
The focus of this paper is to explore the benefits of allowing cooperation among the transmitters and the receivers to enable joint transmission and reception of the messages.  
 
Consider a scenario where the transmitters are connected to each other through a backhaul link. The transmitters could exchange the messages with each other through the backhaul so that multiple transmitters jointly transmit information to the receivers. We capture the cost of cooperation through a number $M_t$, called the {\em transmit cooperation order}, which denotes the number of transmitters having access to each message. We refer to this channel as the interference channel with Coordinated Multi-Point transmission (CoMP) transmission. Note that this model fits well in the context of a cellular downlink with a high-speed fiber-optic or microwave backhaul connecting the base stations, and the acronym CoMP is widely used by the fourth generation cellular standards \cite{marsch2011coordinated}.
 
Similarly, consider a scenario where the receivers are connected through a backhaul and the decoder of a message has the knowledge of the signals received at multiple receivers. The number $M_r$, referred to as the {\em receive cooperation order}, represents the number of receivers that jointly decode each message. We refer to this channel as the interference channel with CoMP reception. This model fits well in the context of the cellular uplink. We could in general consider the interference channel with both CoMP transmission and CoMP reception. For simplicity, we use the term {\em CoMP channel} to denote the interference channel with CoMP transmission and CoMP reception. In \cite{wang2011interference}, a potential application for studying such a channel is presented. Consider a three-hop wireless network scenario with the interference channel at the center forming a bottleneck. Each transmitter has access to multiple message sources, where the decoder of each message has access to the signals received at multiple receivers.  
 
Our objective in this paper is to characterize the Degrees of Freedom (DoF) of the CoMP channel as a function of $K$, $M_t$ and $M_r$. The DoF, also known as the multiplexing gain and the pre-log factor, can be interpreted as the total number of interference-free channels that can be created using the original channel. If a channel has DoF equal to $d$, then the sum capacity scales with SNR as $\Omega(d\log\text{SNR})$. Following up on the breakthrough papers \cite{maddah2008communication,jafar2008xchannel,cadambe2008interference}, in which the DoF of the X-Channel and the $K$-user interference channel are characterized, recent efforts \cite{cadambe2009interference,cadambe2009degrees,gou2010degrees,motahari2009real} have characterized the DoF of many other wireless channels. The proof techniques developed in the above papers are inadequate in characterizing the DoF of the CoMP channel. In references \cite{yetis2010feasibility,bresler2011settling,razaviyayn2011degrees}, tools from Algebraic Geometry are used to determine the achievable DoF using beamforming techniques in MIMO interference channels (without cooperation). These concepts from algebraic geometry play a central role in determining the DoF of the CoMP channel. Specifically, we exploit the notion of algebraic independence of rational functions, and the Jacobian criterion for verifying the algebraic independence, to determine the achievable schemes. 
\subsection{Organization} 
The rest of the paper is organized is as follows. In Section~\ref{sec:model}, we introduce the channel model. In Section~\ref{sec:summary}, we summarize the related work.  In Section~\ref{sec:comp-ob}, we prove an outer bound on the DoF of the CoMP channel. In Section~\ref{sec:app-ag}, we summarize the necessary concepts from algebraic geometry, and prove a useful technical lemma. In Section~\ref{sec:comp-fulldof}, we derive conditions on the transmit and receive cooperation orders such that the DoF of the CoMP channel is equal to $K$. In Sections~\ref{sec:comp-tx}, \ref{sec:comp-tx-mk2} and \ref{sec:comp-tx-mk}, we present achievable schemes for the interference channel with CoMP transmission. In Section~\ref{sec:concl}, we provide some concluding remarks.  
\subsection{Notation} 
We use the following notation. For deterministic objects, we use lowercase letters for scalars, lowercase letters in bold font for vectors, and uppercase letters in bold font for matrices. For example, we use $h$ to denote a deterministic scalar and $\bf{h}$ to denote a deterministic vector, and $\bf{H}$ to denote a deterministic matrix. For random objects, we use uppercase letters for scalars, and underlined uppercase letters for vectors. Random objects with superscripts denote sequences of the random objects in time. For example, we use $X$ to denote a random scalar, $\underline{x}$ to denote a random vector, and $X^n$ and $\underline{x}^n$ to denote the sequences of length $n$ of the random scalars and vectors, respectively.

Given the matrix $\bf{H}$ and the ordered sets $\sA, \sB$, we use $\bH(\sA,\sB)$ to denote the $|\sA| \times |\sB|$ submatrix of $\bH$ obtained by retaining rows indexed by $\sA$ and columns indexed by $\sB$.  We use $\sK$ to denote the set $\sK = \{1,2,\cdots,K\}$, where the number $K$ will be obvious from the context. For any $m \leq K$, we use $k \uparrow m$ and $k \downarrow m$ to denote the sets 
\[
\begin{split}
k \uparrow m & = \{k,k+1,k+2,\cdots,k+m-1\} \\
k \downarrow m & = \{k,k-1,k-2,\cdots,k-m+1\}.
\end{split}
\]
The indices are taken modulo $K$ such that $k \uparrow m, k \downarrow m \subseteq \sK$. Observe that for any two indices $i,j$ and $m \leq K$, $i \in j \uparrow m$ is true if and only if $j \in i \downarrow m$. 
\section{Channel Model} \label{sec:model} 

Consider transmitting $K$ independent messages over the SISO Gaussian interference channel with $K$ transmitters and $K$ receivers:
\be{eq:comp-ic-siso}
Y_{i} = \sum_{j = 1}^{K}h_{ij}X_{j} + Z_{i}, \, \forall i \in \sK
\ee
with an average transmit power constraint of $P$ at each transmitter. In fact, we consider $L$ such parallel Gaussian interference channels, providing  the encoders and decoders an opportunity to jointly encode and jointly decode the messages over the $L$ parallel channels. We can combine the $L$ parallel channels and express them together as one MIMO Gaussian interference channel
\be{eq:comp-ic}
\uy_{i} = \sum_{j = 1}^{K}\bH_{ij}\ux_{j} + \uz_{i}, \, \forall i \in \sK
\ee
such that the channel transfer matrices are square and diagonal. The channel transfer matrix $\bH_{ij}$ is given by
\[
\bH_{ij} = \mat{h_{ij}(1) \\ & h_{ij}(2) \\ & & \ddots  \\ & & & h_{ij}(L)}
\]
where $h_{ij}(\ell) \in \mathbb{C}$ denotes the complex channel coefficient from transmitter $j$ to receiver $i$ in the $\ell$th parallel channel. The reason for considering the parallel channels will be clear at a later stage. 
\subsection{CoMP Model} \label{sec:comp-model}
 
We consider transmitting $K$ independent messages over the channel \eqref{eq:comp-ic} with message $W_{k}$ intended for receiver $k$. In the CoMP setup, we assume that the transmitters cooperatively transmit these messages to the receivers. For each $k \in \sK$, the message $W_k$ is transmitted jointly by the transmitters from the {\em transmit set} $\sT_k$ given by  
\be{eq:tx-sets} 
\sT_k = k \uparrow M_t = \{k,k+1,\cdots,k+M_t-1\}. 
\ee 
The number $M_t$, referred to as the {\em transmit cooperation order}, controls the level of cooperation allowed. Observe that this model allows for a natural transition from the interference channel with no cooperation to the broadcast channel with perfect cooperation; these two extreme cases can be recovered by setting $M_{t}=1$ and $M_{t}=K$ respectively.  
 
We now allow for receive cooperation by letting multiple receivers jointly decode messages. For each $k \in \sK$, we define the {\em receive set} $\sR_k$ as 
\be{eq:rx-sets} 
\sR_k = k \uparrow M_r = \{k,k+1,\cdots,k+M_r-1\}. 
\ee 
The receivers in the receive set $\sR_k$ jointly decode the message $W_k$, i.e., the decoder of message $W_{k}$ has access to the signals $\{y_i: i \in \sR_k\}$. The number $M_r$ is referred to as the {\em receive cooperation order}. Observe that our model covers the interference, broadcast, multiple-access, and point-to-point channels as special cases: 
\begin{enumerate} 
\item $(M_{t},M_{r}) = (1,1)$: No cooperation is allowed either at the transmitters or at the receivers, and hence we obtain the $K$-user interference channel. 
\item $(M_{t},M_{r}) = (K,1)$: All the $K$ transmitters cooperate to jointly transmit the $K$ messages, and hence we obtain a $K$-user broadcast channel. 
\item $(M_{t},M_{r}) = (1,K)$: All the $K$ receivers cooperate to jointly decode the $K$ messages, and hence we obtain a $K$-user multiple access channel.
\item $(M_{t},M_{r}) = (K,K)$: We have perfect cooperation at both the transmitters and the receivers, and hence we obtain a point-to-point MIMO channel with $K$ transmit antennas and $K$ receive antennas.  
\end{enumerate} 
Thus the CoMP channel is specified by the parameters $K,M_{t},M_{r}$ and $L$, denoting the number of users, transmit cooperation order, receive cooperation order, and the number of parallel channels, respectively.

\subsection{Achievable Scheme} 
For each $k \in \sK$, the message $W_k$ is transmitted jointly by the transmitters in the transmit set $\sT_{k}$ and is jointly received by the receivers in the receive set $\sR_{k}$. A communication scheme consists of $K$ encoders and $K$ decoders. Each transmitter is associated with an encoder and each receiver is associated with a decoder. We consider the block coding schemes with $n$ denoting the block length. For a fixed rate tuple $(R_{1},R_{2},\cdots,R_{K}) \in \mathbb{R}_{+}^{K}$ and a block length $n \geq 1$, the message $W_{k}$ is selected from the set $\sW_{k} = \{1,2,\cdots,2^{nR_{k}}\}$. For each $j \in \sK$, the encoder at transmitter $j$ takes the available messages $W_{k}: j \in \sT_{k}$ as inputs and outputs the signal $\ux_{j}^{n}$
\[
\ux_{j}^{n}: \prod_{k: j \in \sT_{k}} \sW_{k} \rightarrow \mathbb{C}^{L \times n} 
\]
satisfying the power constraint 
\[
\mathbb{E}||\ux_{j}^{n}||^{2} \leq nLP.
\] 
For each $k \in \sK$, the decoder of message $W_{k}$ takes the available received signals $\uy_{i}^{n}: i \in \sR_{k}$ as inputs and and reconstructs the message $W_{k}$
\[
\hat{W}_{k}: \mathbb{C}^{nL|\sR_{k}|} \rightarrow \sW_{k}. 
\]
Assuming that the messages are independent and uniformly distributed, any communication scheme is associated with a probability of error $e_n$, defined as $\max_k \textrm{Pr}[\hat{W}_{k} \neq W_{k}]$. A rate tuple $(R_{1},R_{2},\cdots,R_{K})$ is said to be achievable if there exists a sequence of block codes such that $e_n \rightarrow 0$ as $n \rightarrow \infty$. The capacity region $\mathcal{C}(P)$ is defined as the closure of the set of achievable rate tuples. The degrees of freedom  (dof) region $\sD$ is defined as the set of tuples $\mb{d} \in \mathbb{R}_{+}^{K}$ satisfying 
\[ 
\mb{w}^{\top}\mb{d} \leq \limsup_{P \rightarrow \infty} \max_{\mb{R}\in\sC(P)}\frac{\mb{w}^{\top}\mb{R}}{\log P} 
\] 
for each weight vector $\mb{w} \in \mathbb{R}_{+}^{K}$. 
\subsection{Degrees of Freedom} 
Let $\dof(K,M_{t},M_{r},L)$ denote the normalized sum DoF of the CoMP channel with a transmit cooperation order of $M_{t}$ and a receive cooperation order of $M_{r}$ normalized by the number of parallel channels $L$. In general, this number can depend on the specific realizations of channel coefficients
\[
h_{ij}(\ell): i,j \in \sK, 1 \leq \ell \leq L.
\]
However, we ignore this dependency because, in all the known cases, the DoF turns out to be the same for all {\em generic channel coefficients}. We refer the reader to Section~\ref{sec:app-ag} for a precise definition of the generic property. 
Let $\dof(K,M_{t},M_{r})$ denote the asymptotic normalized sum DoF, i.e.,
\[
\dof(K,M_{t},M_{r}) = \limsup_{L \rightarrow \infty} \dof(K,M_{t},M_{r},L).
\]
We say that the DoF is independent of the number of parallel channels $L$ and is equal to some number $d_{\Sigma}$ if and only if $\dof(K,M_{t},M_{r},L) = d_{\Sigma}$ for all $L \geq 1$.
\section{Related Work} \label{sec:summary}  
CoMP transmission (also known as network-MIMO, virtual-MIMO and multi-cell-MIMO) has been identified as one of the study items for fourth generation cellular systems such as LTE-Advanced. There has been considerable interest in devising practical cooperative schemes that improve on uncoordinated schemes, and in estimating the tradeoff between the performance benefits and the additional overhead due to cooperation \cite{annapureddy2010coordinated,gesbertmulti,marsch2011coordinated}. Also, we note that CoMP transmission and reception is just one of the many possible ways for partial transmitter and receiver cooperation in the interference channel. In \cite{prabhakaran2009source,prabhakaran2009destination}, it is assumed that the nodes can both transmit and receive in full-duplex. In \cite{wang2010transmitter,wang2009receiver}, the presence of noise-free finite-capacity links between the transmitter nodes or the receiver nodes is assumed. In \cite{gollakota2009interference}, the receivers are allowed to exchange the decoded messages over a backhaul link to enable interference cancelation. 
 
Special cases of the the CoMP channel have been studied in the past under different names such as  cognitive interference channel \cite{lapidoth2007cognitive,cadambe2007arXiv,devroye06cognitive,devroye07mimox,jovicic2009cognitive}, interference channel with local or partial side-information \cite{lapidoth2008linear,wigger2009receivers}, interference channel with clustered decoding \cite{wang2011interference}, or a combination of thereof \cite{lapidoth2009cognitive}.  However, the DoF of the CoMP channel has not been determined except in some special cases:
\begin{enumerate} 
\item $(M_{t},M_{r}) = (K,1)$: With perfect cooperation at the transmitters, we see that each parallel channel is equivalent to the $K$-user MISO broadcast channel with $K$ transmit antennas. Therefore, we obtain that the DoF is independent of $L$ and is equal to $K$ \cite{viswanath2003sum, vishwanath2003duality,yu2004sum}. 
\item $(M_{t},M_{r}) = (1,K)$: With perfect cooperation at the receivers, we see that each parallel channel is equivalent to the $K$-user SIMO multiple access channel with $K$ receive antennas. Therefore, we obtain that the DoF is independent of $L$ and is equal to $K$ \cite{tse2004diversity}. 
\item $(M_{t},M_{r}) = (K,K)$: With perfect cooperation at the transmitters and at the receivers, we see that each parallel channel is equivalent to the point-to-point MIMO channel with $K$ transmit antennas and $K$ receive antennas.  Therefore, we obtain that the DoF is independent of $L$ and is equal to $K$ \cite{telatar1999capacity, zheng2003diversity}.
\item $(M_{t},M_{r}) = (K-1,1)$ or $(1,K-1)$: For the case where $M_t=K-1$, and $M_r=1$, each message is transmitted jointly using $K-1$ transmit antennas, hence, a zero-forcing beam vector can be used to perfectly null out the interference at $K-1$ receivers. By only scheduling $K-1$ users, it is clear that a sum DoF of $K-1$ can be achieved per each parallel channel. The converse follows easily from Theorem $1$ in \cite{lapidoth2007cognitive}. It is easy to see that similar arguments with receive beamforming hold true when $M_{t} = 1$ and $M_{r} = K-1$. Therefore, we obtain that the DoF is independent of $L$ and is equal to $K-1$.
\item $(M_{t},M_{r}) = (1,1)$: With no cooperation at the transmitter side or the receiver side, we see that each parallel channel is equivalent to the $K$-user Gaussian interference channel. In \cite{cadambe2008interference}, Cadambe and Jafar exploited the channel diversity obtained by considering the parallel channels and proposed a scheme that achieves $K/2$ DoF in an asymptotic fashion. It was already known that the DoF is upper-bounded by $K/2$ \cite{madsen05multiplexing}. The Cadambe-Jafar achievable scheme is a linear beamforming scheme that operates on $L$-parallel Gaussian interference channels simultaneously to create $d$ interference-free channels per user such that $d \rightarrow L/2$ as $L \rightarrow \infty$, thus proving that
\[
\dof(K,1,1) = \lim_{L \rightarrow \infty} Kd/L = K/2.
\]

\end{enumerate} 
To summarize, we know the following results:
\[
\dof(K,M_{t},M_{r}) = \begin{dcases}
K/2 & (M_{t},M_{r}) = 1 \\
K-1 & (M_{t},M_{r}) = (K-1,1) \text{ or } (1,K-1) \\
K & \max(M_{t},M_{r}) = K.
\end{dcases}
\]
\section{Outer Bounds} \label{sec:comp-ob} 
In this section, we derive an outer bound on the DoF as function of $K,M_{t}$ and $M_{r}$. First, we present an outer bound on the DoF region of the CoMP channel with arbitrary transmit and receive sets, i.e., without explicitly using the structure of the transmit sets \eqref{eq:tx-sets}  and the receive sets \eqref{eq:rx-sets}.
\subsection{Outer Bound on DoF Region} \label{sec:comp-ob-region} 
\begin{theorem} \label{thm:comp-dof-ob-region} 
Any point $(d_1,d_2,\cdots,d_K)$ in the normalized (by the number of parallel channels) DoF region of the CoMP channel with generic channel coefficients satisfies the inequalities: 
\be{eq:comp-dof-ob-region} 
\begin{split} 
 \sum_{k: \sT_{k} \subseteq \sA \text{ or } \sR_{k} \subseteq \sB} d_{k} \leq \max(|\sA|,|\sB|),  \forall \sA, \sB \subseteq \sK.
\end{split} 
\ee 
\end{theorem} 
\begin{IEEEproof} 
Without any loss of generality, we can assume $|\sA| = |\sB|$. Otherwise, the smaller set can be blown up to add more terms on the L.H.S. of \eqref{eq:comp-dof-ob-region} without affecting the R.H.S., resulting in an inequality that is stricter than what we need to prove. Now, the objective is to show that  
\be{eq:ob-obj} 
\sum_{k: \sT_{k} \subseteq \sA \text{ or } \sR_{k} \subseteq \sB} d_{k} \leq |\sB|. 
\ee 
Define the subsets 
\[
\begin{split} 
\sW_t & \ = \{W_k: \sT_k \subseteq \sA\} \\ 
\sW_r & \ = \{W_k: \sR_k \subseteq \sB, \sT_k \not\subseteq \sA\} 
\end{split} 
\]
and $\sW_f$ as the set of free messages that do not appear in either of the sets $\sW_r$ and $\sW_t$.  The proof idea is to start with the signals received by the receivers $\sB$, and show that the messages $\sW_t$ and $\sW_r$ can be decoded using these $|\sB|$ received signals with $\sW_f$ as side-information. For any given subset $\sS \subseteq \sK$, we use the notation $\ux_{\sS}$ to denote the vector made up of the signals transmitted by the transmitters in the set $\sS$, with a similar notation used for $\uy_{\sS}$ and $\uz_{\sS}$.

For each $k$, using Fano's inequality and the definition of the receive set $\sR_k$, we have that any reliable communication scheme must satisfy
\[ 
\Ent{W_k|\uy_{\sR_k}^n} \leq n\epsilon_n 
\]
where $\epsilon_n \rightarrow 0$, as $n \rightarrow \infty$. Therefore, we immediately have 
\be{eq:wr} 
\Ent{\sW_r|\uy_{\sB}^n} \leq \sum_{k: \sR_k \in \sB}\Ent{W_k|\uy_{\sB}^n} \leq |\sW_r|n\epsilon_n
\ee 
i.e., the messages $\sW_r$ can be decoded by the receivers $\sB$.  Similarly, the messages $\sW_t$ can be decoded using all the received signals: 
\[ 
\Ent{\sW_t|\uy_{\sK}^n,\sW_r,\sW_f} \leq \Ent{\sW_t|\uy_{\sK}^n}\leq |\sW_t|n\epsilon_n. 
\] 
But, we need to show that the messages $\sW_t$ can also be decoded by the receivers $\sB$ with $\sW_f$ as side-information. We do so by arguing that the signal contribution in $\uy_{\sK}^n$ can be reconstructed using $\sW_f, \sW_r$ and $\uy_{\sB}^n$: 
\[ 
\begin{split} 
& \Ent{\sW_t|\uy_{\sB}^n, \sW_r, \sW_f}  \\ 
& ~~~ \leq \Ent{\sW_t|\uy_{\sB}^n, \sW_r, \sW_f}  - \Ent{\sW_t|\uy_{\sK}^n, \sW_r, \sW_f}  + |\sW_t|n\epsilon_n \\  
& ~~~ = \mi{\sW_t; \uy_{\sB^c}^n|\uy_{\sB}^n,\sW_f,\sW_r} + |\sW_t|n\epsilon_n\\ 
& ~~~ = \ent{\uy_{\sB^c}^n|\uy_{\sB}^n,\sW_f,\sW_r } - \ent{\uy_{\sB^c}^n|\sW_f,\sW_r,\sW_t} + |\sW_t|n\epsilon_n\\ 
& ~~~ = \ent{\uy_{\sB^c}^n|\uy_{\sB}^n,\sW_f,\sW_r } - \ent{\uz_{\sB^c}^n} + |\sW_t|n\epsilon_n \\ 
& ~~~ \leq \ent{\uy_{\sB^c}^n|\uy_{\sB}^n,\ux_{\sA^c}^n } - \ent{\uz_{\sB^c}^n} + |\sW_t|n\epsilon_n . 
\end{split} 
\] 
Observe that, over each symbol, we have 
\[ 
\begin{split} 
\uy_{\sB^c} & \ = \bH(\sB^c,\sA)\ux_{\sA} + \bH(\sB^c,\sA^{c})\ux_{\sA^c} +\uz_{\sB^c}\\ 
\uy_{\sB} & \ = \bH(\sB,\sA)\ux_{\sA} + \bH(\sB,\sA^{c})\ux_{\sA^c} + \uz_{\sB} 
\end{split} 
\] 
where we used $\bH$ to denote the $KL \times KL$ channel transfer matrix from all the $K$ transmitters to the $K$ receivers, i.e.,
\[
\bH = \mat{\bH_{11} & \cdots & \bH_{1K} \\ \vdots & \ddots & \vdots \\ \bH_{K1} & \cdots & \bH_{KK}}
\]
and $\bH(\sB,\sA)$ to denote the $|\sB|L \times |\sA|L$ channel transfer matrix from transmitters $\sA$ to the receivers $\sB$, and $\bH(\sB^{c},\sA^{c})$, $\bH(\sB,\sA^{c})$ and $\bH(\sB^{c},\sA)$ to denote appropriate submatrices.  For generic channel coefficients, since we assumed that $|\sA| = |\sB|$, the matrix $\bH(\sB,\sA)$ is invertible, and hence we have 
\[ 
\begin{split} 
\tilde{\uz} & = \uy_{\sB^c}-\bH(\sB^c,\sA)\ux_{\sA^c}\\ 
& ~~~~~  - \bH(\sB^c,\sA)\bH(\sB,\sA)^{-1}\lt(\uy_{\sB}-\bH(\sB,\sA)\ux_{\sA^c} \rt) \\ 
& = \uz_{\sB^c} - \bH(\sB^c,\sA)\bH(\sB,\sA)^{-1}\uz_{\sB}. 
\end{split} 
\] 
Thus, we get 
\[
\begin{split} 
\Ent{\sW_t|\uy_{\sB}^n, \sW_r, \sW_f}\leq & \ \ent{\tilde{\uz}^n} - \ent{\uz_{\sB^c}^n} + |\sW_t|n\epsilon_n. 
\end{split} 
\] 
Therefore, we have 
\[
\begin{split} 
\Ent{\sW_r,\sW_t} & \ \leq \Ent{\sW_r,\sW_t|\sW_f} \\ 
& \ = \mi{\sW_r,\sW_t;\uy_\sB^n|\sW_f} +  \Ent{\sW_r,\sW_t|\uy^n_\sB,\sW_f}\\ 
& \ = \ent{\uy_\sB^n|\sW_f} - \ent{\uz_\sB^n} +  H(\sW_r|\uy^n_\sB,\sW_f) + H(\sW_t|\sW_r,\uy^n_\sB,\sW_f)\\ 
& \ \leq \ent{\uy_\sB^n} - \ent{\uz_\sB^n} + \ent{\tilde{\uz}^n} - \ent{\uz_{\sB^c}^n} + (|\sW_t| + |\sW_r|)n\epsilon_n. 
\end{split} 
\]
Observe that all the terms, except for $ \ent{\uy_\sB^n}$, are independent of the power constraint $P$. Furthermore, the sequence $\uy_\sB^n$ denotes a vector of length $n|\sB|L$. Therefore, there must exist a constant $c$ that may depend on the channel coefficients, but is independent of the power constraint $P$ and the block length $n$ such that
\[
\Ent{\sW_r,\sW_t} \leq n|\sB|L\log P + nc + (|\sW_t| + |\sW_r|)n\epsilon_n.
\]
Therefore, any achievable rate tuple $(R_{1},R_{2},\cdots,R_{K})$ must satisfy
\[
\sum_{k: \sT_{k} \subseteq \sA \text{ or } \sR_{k} \subseteq \sB}R_{k} \leq |\sB|L\log P + c
\]
which immediately implies that any achievable DoF vector (normalized by the number of parallel channels $L$) must satisfy \eqref{eq:ob-obj}. 
\end{IEEEproof} 
\subsection{Outer Bound on Sum DoF} \label{sec:comp-ob-sumdof}
We use Theorem~\ref{thm:comp-dof-ob-region} to obtain an outer bound on $\dof(K,M_{t},M_{r},L)$. Observe that an obvious outer bound given by
\[
\dof(K,M_{t},M_{r},L) \leq K
\]
can be obtained by setting $\sA = \sB = \sK$. The following theorem provides a nontrivial outer bound when $M_{t} + M_{r} \leq K$.
\begin{theorem} \label{thm:comp-ob-sumdof} 
The (normalized sum) DoF of the CoMP channel with generic channel coefficients satisfies
\[
\dof(K,M_t,M_r,L) \leq \left\lceil\frac{K+M_t+M_r-2}{2}\right\rceil. 
\]
When $K+M_t+M_r$ is odd, the above outer bound can be improved to obtain
\[
\dof(K,M_t,M_r,L) \leq \frac{K}{K-1}\frac{K+M_t+M_r-3}{2}.
\]
\end{theorem} 
\begin{IEEEproof}
First, observe that the stated outer bounds are weak compared to the obvious outer bound $\dof(K,M_{t},M_{r},L) \leq K$ if $M_{t} + M_{r} \geq K+1$. Therefore, we assume that $M_{t} + M_{r} \leq K$ in proving the theorem. The best outer bound on $\dof(K,M_t,M_r)$ that we can obtain using Theorem~\ref{thm:comp-dof-ob-region} is obtained by solving the linear program 
\[  
\max_{(d_1,\cdots,d_K)}d_1 + d_2 \cdots + d_K 
\]
subject to the constraints \eqref{eq:comp-dof-ob-region}, given by
\[
\sum_{k \in \sK: \sT_{k} \subseteq \sA \text{ or } \sR_{k} \subseteq \sB}d_{k} \leq r
\]
for every $\sA, \sB \subseteq \sK$ such that $|\sA| = |\sB| = r$.
Since the the transmit sets \eqref{eq:tx-sets} and receive sets \eqref{eq:rx-sets} are symmetric across the transmitter and receiver indices, by appropriately averaging the above upper bound by fixing $r$, and rotating the sets $\sA$ and $\sB$, we obtain the following upper bound on the normalized sum DoF:
\[
\dof(K,M_t,M_r,L) \leq \frac{Kr}{|k \in \sK: \sT_{k} \subseteq \sA \text{ or } \sR_{k} \subseteq \sB|}.
\]
Therefore, the objective is to choose the sets $\sA$ and $\sB$ so that the ratio on the R.H.S. of the above inequality is minimized. Since $\sT_{k} = k \uparrow M_{t}$, and $|\sA| = r$, we have that 
\[
|k \in \sK: \sT_{k} \subseteq \sA| = (r - M_{t}+1)_{+}.
\]
Similarly, we have that 
\[
|k \in \sK: \sR_{k} \subseteq \sB| = (r - M_{r}+1)_{+}.
\]
where $(x)_{+}$ is defined as $\max(x,0)$. Clearly, $r$ must satisfy $r \leq K$. It can be easily argued that, without any loss of generality, we can also restrict $r$ so that $r - M_{t}+1 \geq 1$ and $r - M_{r}+1 \geq 1$ and $2r - M_{t} - M_{r} + 2 \leq K$. For any such value of $r$, we can choose the sets $\sA$ and $\sB$  to be
\[ 
\begin{split} 
\sA & = \{1,2,\cdots,r\}\\ 
\sB & = \{r-M_t+2,r-M_t+3,\cdots,2r-M_t+1\}. 
\end{split} 
\] 
so that the sets $\{k \in \sK: \sT_{k} \subseteq \sA\}$ and $\{k \in \sK: \sR_{k} \subseteq \sB\}$ do not intersect. This results in the outer bound
\[
\dof(K,M_t,M_r,L) \leq \frac{Kr}{2r - M_{t} - M_{r} + 2}.
\]
To obtain the best possible outer bound, it is clear that we should choose $r$ to be as high as possible while satisfying the conditions $2r - M_{t} - M_{r}+2 \leq K$ and $r \leq K$. When $K + M_{t} + M_{r}$ is even, the best is to set
\[
r = \frac{K+M_t+M_r-2}{2}
\]
resulting in the required outer bound $\dof(K,M_t,M_r,L) \leq r$. When $K + M_{t} + M_{r}$ is odd, the best is to set
\[
r = \frac{K+M_t+M_r - 3}{2}
\]
resulting in the required outer bound $\dof(K,M_{t},M_{r},L) \leq Kr/(K-1)$.
\end{IEEEproof}

We now prove that the outer bound in Theorem~\ref{thm:comp-ob-sumdof} is achievable in some special cases. The achievability proofs depend heavily on techniques from algebraic geometry. We first review these techniques and then proceed to prove the achievability results. 
\section{Mathematical Preliminaries} \label{sec:app-ag} 
 
In this appendix, we present some results in algebraic geometry that are essential in proving the achievability results. We start by recalling some basic terminology in algebraic geometry. We refer the reader to the book \cite{cox2007ideals} for an excellent introduction. 
\subsection{Varities and Ideals} 
Let $\mathbb{C}[t_{1},t_{2},\cdots,t_{n}]$ and $\mathbb{C}(t_{1},t_{2},\cdots,t_{n})$ denote the set of multivariate polynomials and rational functions, respectively, in the variables $t_{1},t_{2},\cdots,t_{n}$. For any polynomials $f_{1},f_{2},\cdots,f_{m} \in \mathbb{C}[t_{1},t_{2},\cdots,t_{n}]$, the {\em affine variety} generated by $f_{1},f_{2},\cdots,f_{m}$ is defined as set of points at which the polynomials vanish: 
\[ 
V(\mathbf{f}) = \{\bt \in \mathbb{C}^{n}: \mathbf{f}(\bt) = \bo\}. 
\] 
Any subset $I \subseteq \mathbb{C}[t_{1},t_{2},\cdots,t_{n}]$ is called an {\em ideal} if it satisfies the three properties 
\begin{itemize} 
\item $0 \in I$. 
\item If $f_{1},f_{2} \in I$, then $f_{1}+f_{2} \in I$. 
\item If $f_{1} \in I$ and $f_{2} \in \mathbb{C}[t_{1},t_{2},\cdots,t_{n}]$, then $f_{1}f_{2} \in I$. 
\end{itemize} 
For any set $\sA \subseteq \mathbb{C}^{n}$, the ideal generated by $\sA$ is defined as 
\[ 
I(\sA) = \{f \in \mathbb{C}[t_{1},t_{2},\cdots,t_{n}]: f(\bt) = 0 ~ \forall \bt \in \sA\}. 
\] 
For any ideal $I$, the affine variety generated by $I$ is defined as 
\[
V(I) = \{\bt \in \mathbb{C}^{n}: f(\bt) = 0 ~ \forall f \in I\}. 
\]
The {\em Zariski topology} on the affine space $\mathbb{C}^{n}$ is obtained by taking the affine varieties as closed sets. For any set $\sA \in \mathbb{C}^{n}$, the Zariski closure $\bar{\sA}$ is defined as 
\[
\bar{\sA} = V(I(\sA)). 
\] 
A set $\sA \subseteq \mathbb{C}^{n}$ is said to be {\em constructible} if it is a finite union of locally closed sets of the form $U \cap Z$ with $U$ closed and $Z$ open. If $\sA \subseteq \mathbb{C}^{n}$ is constructible and $\bar{\sA} = \mathbb{C}^{n}$, then $\sA$ must be dense in $\mathbb{C}^{n}$, i.e., $\sA^{c} \subseteq W$ for some non-trivial variety $W \subsetneq \mathbb{C}^{n}$.
\subsection{Algebraic Independence and Jacobian Criterion} 
The rational functions $f_1,f_2\cdots,f_m \in \mathbb{C}(t_1,t_2,\cdots,t_n)$ are called algebraically dependent (over $\mathbb{C}$) if there exists a nonzero polynomial $F \in \mathbb{C}[s_1,s_{2}\cdots,s_m]$ such that $F(f_1,f_2,\cdots,f_m)  = 0$. If there exists no such annihilating polynomial $F$, then $f_1,f_2,\cdots,f_m$ are algebraically independent. 
\begin{lemma}[Theorem~3 on page 135 of \cite{hodge1968methods}] \label{lemma:app-ag-jacob} 
The rational functions $f_1,f_2\cdots,f_m \in \mathbb{C}(t_1,t_2,\cdots,t_n)$ are algebraically independent if and only if the Jacobian matrix 
\be{eq:jacob-s-t}         
\bJ_{f} =\lt(\frac{\partial f_{i}}{\partial t_{j}}\rt)_{1\leq i\leq m, 1 \leq j \leq n} 
\ee 
has full row rank equal to $m$.
\end{lemma} 

The Jacobian matrix is a function of the variables $t_{1},t_{2},\cdots,t_{n}$, and hence the Jacobian matrix can have different ranks at different points $\bt \in \mathbb{C}^{n}$.  The above lemma refers to the {\em structural rank} of the Jacobian matrix which is equal to $m$ if and only if there exists at least one realization $\bt \in \mathbb{C}^{n}$ where the Jacobian matrix has full row rank. 
\subsection{Dominant Maps and Generic Properties} 
A polynomial map $\mathbf{f}: \mathbb{C}^{n} \rightarrow \mathbb{C}^{m}$ is said to be {\em dominant} if the Zariski closure of the image $\mathbf{f}(\mathbb{C}^{n})$ is equal to $\mathbb{C}^{m}$. The image of a polynomial map is constructible. Therefore, the image of a dominant polynomial map is dense, i.e., the complement of $\mathbf{f}(\mathbb{C}^{n})$ is contained in a non-trivial variety $W \subsetneq \mathbb{C}^{m}$. The implication of this is that the system of polynomial equations 
\be{eq:poly-s-t} 
\begin{split} 
s_{1}= & \ f_{1}(t_{1},t_{2},\cdots,t_{n}) \\ 
s_{2}= & \ f_{2}(t_{1},t_{2},\cdots,t_{n}) \\ 
~~ \vdots \\ 
s_{m} = & \ f_{m}(t_{1},t_{2},\cdots,t_{n}) 
\end{split} 
\ee 
has a solution $\bt \in \mathbb{C}^{n}$ for generic $\bs$, where the notion of a generic property is defined below.  
\begin{definition} 
A property is said to true for generic $\bs \in \mathbb{C}^{m}$ if the property holds true for all $\bs \in \mathbb{C}^{m}$ except on a non-trivial affine variety $W \subsetneq \mathbb{C}^{m}$. Such a property is said be a generic property. 
\end{definition} 
 
For example, a generic square matrix $\bA$ has full rank because $\bA$ is rank deficient only when it lies on the affine variety generated by the polynomial $f(\bA) = \det{\bA}$. If the variables are generated randomly according to a continuous joint distribution, then any generic property holds true with probability $1$. 
 
Observe that the Zariski closure of the image $\mathbf{f}(\mathbb{C}^{n})$ is equal to $\mathbb{C}^{m}$ if and only if the ideal $I$ generated by the image set is equal to $\{0\}$. Since $I$ is equal to the set of annihilating polynomials 
\[
\begin{split} 
I & = \{F \in \mathbb{C}[s_{1},s_{2},\cdots,s_{m}]: F(\bs) = 0 ~ \forall \bs \in \mathbf{f}(\mathbb{C}^{n})\} \\ 
& = \{F \in \mathbb{C}[s_{1},s_{2},\cdots,s_{m}]: F(f_{1},f_{2},\cdots,f_{m}) = 0\}, 
\end{split} 
\]
the map $\mathbf{f}$ is dominant if and only if the polynomials $f_{1},f_{2},\cdots,f_{m}$ are algebraically independent. Thus we obtain the following lemma. 
\begin{lemma} \label{lemma:app-ag-dominant-map} 
The system of polynomial equations \eqref{eq:poly-s-t} admits a solution for a generic $\bs \in \mathbb{C}^{m}$ if and only if the polynomials $f_{1},f_{2},\cdots,f_{m}$ are algebraically independent, i.e., if and only if the Jacobian matrix \eqref{eq:jacob-s-t} has full row rank. 
\end{lemma} 
\subsection{A Lemma on Full-Rankness of Certain Random Matrix} 
Let $\bt \in \mathbb{C}^{n}$ be a set of original variables, and let $\bs \in \mathbb{C}^{m}$ be a set of derived variables obtained through polynomial transformation $\bs = \mathbf{f}(\bt)$ for some rational map $\mathbf{f}$. Suppose we generate $p$ instances of $\bt$ 
\be{eq:t-q} 
\bt(1), \bt(2), \cdots, \bt(p) 
\ee
and  the corresponding $p$ instances of $\bs$ 
\[
\bs(1), \bs(2), \cdots, \bs(p) 
\]
and generate the $p \times q$ matrix 
\[
\bM = \mat{ 
\bs(1)^{\ba_{1}} & \bs(1)^{\ba_{2}} & \cdots & \bs(1)^{\ba_{q}} \\ 
\bs(2)^{\ba_{1}} & \bs(2)^{\ba_{2}} & \cdots & \bs(2)^{\ba_{q}} \\ 
\vdots & \vdots & \ddots & \vdots \\ 
\bs(p)^{\ba_{1}} & \bs(p)^{\ba_{2}} & \cdots & \bs(p)^{\ba_{q}} 
} 
\]
for some exponent vectors $\ba_{1},\ba_{2},\cdots,\ba_{q} \in \mathbb{Z}_{+}^{m}$ and $p \geq q$. We are interested in determining the set of variables \eqref{eq:t-q} such that the matrix $\bM$ has full column rank. If there exists an annihilating polynomial $F \in \mathbb{C}[s_{1},s_{2},\cdots,s_{m}]$ of the form 
\be{eq:F-as} 
F(\bs) = \sum_{i = 1}^{q} c_{i}\bs^{\ba_{i}} 
\ee 
such that $F(f_{1},f_{2},\cdots,f_{m}) = 0$, then the matrix $\bM$ satisfies $\bM\bc = \bo$, and hence the matrix $\bM$ does not have full column rank for any realizations of the variables \eqref{eq:t-q}. Interestingly, even the converse holds true.  
\begin{lemma} \label{lemma:fullrank}
The matrix $\bM$ has full column rank for generic realizations of the variables \eqref{eq:t-q} if and only if there does not exist an annihilating polynomial $F$ of the form \eqref{eq:F-as} satisfying $F(f_{1},f_{2},\cdots,f_{m}) = 0$. 
\end{lemma}  
The proof is relegated to Appendix~\ref{sec:app-fullrank}. If the rational functions $f_{1},f_{2},\cdots,f_{m}$ are algebraically independent, then there cannot exist an annihilating polynomial $F$ (of any form) satisfying $F(f_{1},f_{2},\cdots,f_{m}) = 0.$ Thus, we immediately have the following corollary.  
\begin{corollary} \label{corr:app-ag-fullrank}
The matrix $\bM$ has full column rank for generic realizations of the variables \eqref{eq:t-q} if the rational functions $f_{1},f_{2},\cdots,f_{m}$ are algebraically independent, i.e., if the Jacobian matrix \eqref{eq:jacob-s-t} has full row rank. 
\end{corollary} 
\section{Full DoF with Partial Cooperation} \label{sec:comp-fulldof}
Recall from Section~\ref{sec:summary} that the DoF of the CoMP channel is equal to $K$ if perfect cooperation is allowed at either the transmitter side or the receiver side, i.e.,
\[
\dof(K,M_{t},M_{r}) = K \textrm{ if } \max(M_{t},M_{r}) = K. 
\]
In this section, we obtain a necessary and sufficient condition on $M_{t}$ and $M_{r}$ such that the DoF is equal to $K$. First, we can obtain some intuition on the condition from the outer bound in Section~\ref{sec:comp-ob}. Observe that Theorem~\ref{thm:comp-ob-sumdof} says that the DoF is strictly less than $K$ whenever $M_{t} + M_{r} \leq K$. We show that the DoF is equal to the maximum value $K$ whenever $M_{t} + M_{r} \geq K+1$.


\begin{theorem} \label{thm:comp-fulldof} 
The DoF of the CoMP channel with generic channel coefficients is independent of $L$, and is equal to $K$, if and only if $M_{t}$ and $M_{r}$ satisfy $M_{t} + M_{r} \geq K+1$; i.e., 
\[ 
\dof(K,M_{t},M_{r}) = K \Leftrightarrow M_{t} + M_{r} \geq K+1. 
\]
\end{theorem} 

The achievable scheme is based on the linear transmit and receive beamforming strategy over each parallel channel. We prove the theorem assuming $L = 1$, and the general case follows by treating each parallel channel separately. Let $\bV$ and $\bU$ be the $K \times K$ matrices representing the transmit and receive beams respectively. The $k^{\textrm{th}}$ column of $\bV$ (resp. $\bU$) represents the beam along which the message $W_{k}$ is transmitted (resp. received). To comply with the physical constraints imposed by the transmit sets \eqref{eq:tx-sets} and the receive sets \eqref{eq:rx-sets}, the matrices $\bV$ and $\bU$ must satisfy 
\be{eq:tx-rx-sets} 
\begin{split} 
& v_{ik} \neq 0 \Rightarrow i \in \sT_{k} = k \uparrow M_t \\ 
& u_{ik} \neq 0 \Rightarrow i \in \sR_{k} = k \uparrow M_r. 
\end{split} 
\ee 
Let $\bH$ denote the $K \times K$ channel transfer matrix. If $M_{t}$ and $M_{r}$ satisfy $M_{t} + M_{r} \geq K+1$, then we prove the existence of $\bV$ and $\bU$ satisfying \eqref{eq:tx-rx-sets}, and 
\be{eq:diag} 
\bU^\top\bH\bV = \bI 
\ee 
for a generic matrix $\bH$. Observe that the above choice for beamfroming matrices $\bV$ and $\bU$ achieves $K$ DoF since they create $K$ interference-free AWGN channels, one per each message, with each channel having a nonzero SNR. Since $\bU$ and $\bV$ are square matrices, it is easy to see that \eqref{eq:diag} is equivalent to
\be{eq:diag-smd} 
\begin{split} 
\bH^{-1} =  & \ \bV\bU^\top. 
\end{split} 
\ee 
Thus, it remains to show that the $\bH^{-1}$ admits the matrix decomposition in \eqref{eq:diag-smd} for a generic $\bH$. We now prove a more general result.
\subsection{Structural Matrix Decomposition}
Observe that the above matrix decomposition problem\eqref{eq:diag-smd}  is similar to the LU decomposition in the sense that we are interested in expressing a matrix $\bA = \bH^{-1}$ as a product of two matrices $\bV$ and $\bU^{\top}$ with structural constraints on $\bV$ and $\bU$. In the case of LU decomposition, we require that both $\bV$ and $\bU$ are lower triangular matrices, whereas in \eqref{eq:diag-smd} we require $\bV$ and $\bU$ to satisfy the structural conditions \eqref{eq:tx-rx-sets}. In this section, we consider the general problem of structural matrix decomposition (SMD) that generalizes both \eqref{eq:tx-rx-sets} and LU decomposition. We need the following definition to formulate the SMD problem.  
\begin{definition}[S-matrix] Given a matrix $\bV$ and a $(0,1)$-matrix $\bar{\bV}$ of the same size, we say that $\bar{\bV}$ is a structural matrix (or S-matrix) of $\bV$ if $\bar{v}_{ij} = 1$ for all $i,j$ such that $v_{ij} \neq 0$. 
\end{definition}
\begin{example} 
Suppose $\bV$ and $\bU$ be transmit and receive beamforming matrices satisfying the conditions \eqref{eq:tx-rx-sets} corresponding to the setting $K = 3$ and $(M_{t},M_{r}) = (2,2)$. Then, the S-matrices of $\bV$ and $\bU$ are given by
\be{eq:comp-k3mt2mr2} 
\bar{\bU} = \bar{\bV} = \mat{1 & 0 & 1\\ 1 & 1 & 0 \\ 0 & 1 & 1}
\ee 
where the ones in the $k$th column of $\bar{\bV}$ correspond to the transmit set $\sT_{k}$, and the ones in the $k$th column of $\bar{\bU}$ correspond to the receive set $\sR_{k}$.
\end{example}

\begin{definition}[SMD] 
Let $\bA$ be a square matrix, and $\bar{\bV}, \bar{\bU}$ be $(0,1)$-matrices of same size. We say that the matrix $\bA$ admits a structural matrix decomposition (SMD) with respect to $\bar{\bV}$ and $\bar{\bU}$ if $\bA$ can be factorized as
\[
\bA = \bV\bU^{\top} 
\] 
with $\bar{\bV}$ and $\bar{\bU}$ being S-matrices of $\bV$ and $\bU$ respectively. 
\end{definition} 

To prove that $\dof(3,2,2) = 3$, we need to show that a generic $3 \times 3$ matrix $\bA$ admits an SMD with respect to $\bar{\bV}$ and $\bar{\bU}$ defined in \eqref{eq:comp-k3mt2mr2}. The LU decomposition can be seen as a special case of the SMD with $\bar{\bV}$ and $\bar{\bU}$ given by 
\be{eq:comp-lu} 
\bar{\bU} = \bar{\bV} = \mat{1 & 0 & 0\\ 1 & 1 & 0 \\ 1 & 1 & 1}. 
\ee
 
We know that a generic matrix $\bA$ admits an LU decomposition, i.e., a generic matrix $\bA$ admits an SMD if $\bar{\bU}$ and $\bar{\bV}$ are given by \eqref{eq:comp-lu}. We shall show that the same holds true even if \eqref{eq:comp-lu} is replaced with \eqref{eq:comp-k3mt2mr2}. The following theorem provides a sufficient condition on $\bar{\bV}$ and $\bar{\bU}$ such that a generic matrix admits an SMD. 
\begin{theorem} \label{thm:smd} 
Suppose the $K \times K$ $(0,1)$-matrices $\bar{\bV}$ and $\bar{\bU}$ satisfy the conditions 
\begin{enumerate} 
\item The diagonal entries of $\bar{\bV}$ and $\bar{\bU}$ are nonzero.
\item The matrix $\bar{\bV} + \bar{\bU}^{\top}$ is a full matrix, i.e., all of its entries are nonzero. 
\end{enumerate} 
Then, a generic $K \times K$ matrix $\bA$ admits an SMD $\bA = \bV\bU^\top$ with respect to the S-matrices $\bar{\bV}$ and $\bar{\bU}$.
\end{theorem} 
\begin{IEEEproof} 
Suppose a matrix $\bA$ admits an SMD  $\bA = \bV\bU^\top$; then the decomposition is not unique since for any full rank diagonal matrix $\Lambda$, we have 
\[
\bA = \bV\bU^\top = \left(\bV\Lambda\right)\left(\bU\Lambda^{-1}\right)^\top. 
\]
To avoid such degeneracy, we set $u_{kk}=1$ for all $k \in \sK$. We now interpret $\bA = \bV\bU^\top$ as a system of polynomial equations 
\be{eq:polys-md} 
a_{ij} = f_{ij}(\bt), \forall i,j \in \sK 
\ee 
where $\bt$ represents those elements of $\bV$ and $\bU$ that can take arbitrary values, i.e., $\bt$ contains the variables
\be{eq:vars} 
\{v_{ij}: \bar{v}_{ij} = 1\} \cup \{u_{ij}: i \neq j \text{ and } \bar{u}_{ij} = 1\}. 
\ee 
Let $N_{v}$ denote the number of variables so that $\bt \in \mathbb{C}^{N_{v}}$. Our objective is show that the system of equations \eqref{eq:polys-md} has a solution $\bt \in \mathbb{C}^{N_{v}}$ for a generic matrix $\bA$. From Lemma~\ref{lemma:app-ag-dominant-map} in Section~\ref{sec:app-ag}, it follows that \eqref{eq:polys-md} admits a solution for generic $\bA$ if and only if the Jacobian matrix $\bJ_{f}$ of the polynomial map 
\[
\mathbf{f}: \mathbb{C}^{N_{v}} \rightarrow \mathbb{C}^{K \times K} 
\]
has full row rank at some point $\bt^{*}$.

We now prove that $\bJ_{f}$ has full row rank, equal to $K^{2}$, by explicitly computing the Jacobian matrix $\bJ_{f}$ at the point $\bt^{*}$ corresponding to $\bU^* = \bV^* = \bI$. Observe that the two conditions in the theorem statement ensure that for every $i,j \in \sK$, either $v_{ij}$ or $u_{ji}$ is a variable. Thus, $N_{v} \geq K^{2}$, which is a necessary condition for the Jacobian matrix to be a fat matrix, and to have full row rank. Observe that $\bJ_{f}$ has full row rank if any $K^{2} \times K^{2}$ submatrix has full rank. We consider the submatrix corresponding to the $K^{2}$ variables $\{t_{ij}: i,j \in \sK\}$ defined such that $t_{ij}$ is equal to either $v_{ij}$ or $u_{ji}$ for each $i,j\in \sK$. Consider the partial derivative 
\[
\begin{split} 
\frac{\partial a_{pq}}{\partial t_{ij}} & \ = \frac{\partial f_{pq}(\bt)}{\partial t_{ij}} \\ 
& \ = \frac{\partial \sum_{\ell=1}^{K} v_{p\ell}u_{q\ell}}{\partial t_{ij}} \\ 
& \ = \sum_{\ell=1}^{K}\frac{\partial  (v_{p\ell}u_{q\ell})}{\partial t_{ij}}.
\end{split} 
\]
Suppose $t_{ij} = v_{ij}$; then we see that 
\[
\begin{split} 
\frac{\partial a_{pq}}{\partial t_{ij}} & \ = \sum_{\ell=1}^{K}\frac{\partial  (v_{p\ell}u_{q\ell})}{\partial t_{ij}} \\ 
& \ = \delta_{pi}u_{qj}^* \\ 
& \ = \delta_{pi}\delta_{qj}
\end{split} 
\]
where $\delta_{ij}$ is the Kronecker delta function, and in the last step we used the fact that the derivative is taken at the point $t^{*}$ corresponding to $\bU^{*} = \bV^{*} = \bI$. We obtain the same even if $t_{ij} = u_{ji}$. Therefore, we get 
\[ 
\frac{\partial a_{pq}}{\partial t_{ij}} = \left\{ \begin{array}{l l} 1 & \textrm{if } (p,q) = (i,j) \\ 0 & \text{otherwise}.\end{array} \right. 
\]
Thus, we see that the submatrix of $\bJ_{f}$ corresponding to the variables $\{t_{ij}\}$ is equal to the identity matrix. Hence from Lemma~\ref{lemma:app-ag-dominant-map} in Section~\ref{sec:app-ag}, we conclude that a solution to \eqref{eq:polys-md} exists for a generic $\bA$.
\end{IEEEproof}
\subsection{Proof of Theorem~\ref{thm:comp-fulldof}}
To complete the proof of Theorem~\ref{thm:comp-fulldof}, we need to show that the conditions of Theorem~\ref{thm:smd} are satisfied when $M_{t} + M_{r} \geq K+1$. Recall from \eqref{eq:tx-rx-sets} that the S-matrices $\bar{\bV}$ and $\bar{\bU}$ of the beamforming matrices $\bV$ and $\bU$ are given by
\[
\begin{split} 
& \bar{v}_{ij} = 1 \Leftrightarrow i \in j \uparrow M_{t} \\ 
& \bar{u}_{ij} = 1 \Leftrightarrow i \in j \uparrow M_{r}.
\end{split} 
\]
Clearly, the diagonal entries of $\bar{\bV}$ and $\bar{\bU}$ are equal to one satisfying the first condition of Theorem~\ref{thm:smd}. Since $M_{t}  +M_{r} \geq K+1$, for any $(i,j)$ either 
\[
i \in j \uparrow M_{t} \Rightarrow \bar{v}_{ij} = 1 
\]
 or 
\[
i \in j \downarrow M_{r} \Rightarrow j \in i \uparrow M_{r} \Rightarrow \bar{u}_{ji} = 1. 
\]
This verifies that the second condition of Theorem~\ref{thm:smd} is also satisfied. Therefore, we see that the matrix $\bH^{-1}$ admits SMD \eqref{eq:diag-smd} for a generic $\bH$. This completes the proof of Theorem~\ref{thm:comp-fulldof}. 
\subsection{Relation to MIMO Interference Channel and Interference Alignment} \label{sec:comp-fulldof-mimo}

The condition $M_{t} + M_{r} \geq K+1$ is similar to the condition obtained in \cite{yetis2010feasibility} for the MIMO interference channel. The MIMO interference channel with $N_{t} = M_{t}$ antennas per transmitter and $N_{r} = M_{r}$ antennas per receiver is similar to the CoMP channel, in the sense that each message is transmitted and received using $M_{t}$ and $M_{r}$ antennas, respectively. The difference is that the messages in the MIMO interference channel have dedicated antennas, whereas the messages in the CoMP channel share antennas to mimic the MIMO interference channel. In \cite{yetis2010feasibility}, Yetis et al. studied the feasibility of transforming the MIMO interference channel into $K$ interference-free channels using transmit and receive beamforming strategies. They used Bernstein's theorem from algebraic geometry to prove that the beams exist if and only if $M_{t} + M_{r} \geq K+1$. 

The common theme that leads to these results in both the cases, i.e., MIMO interference channel and CoMP channel, is interference alignment.  It is easy to see interference alignment in action in the special case $M_{t} = K-1$ and $M_{r} = 2$ where each decoder has access to two received signals. Out of these two dimensions, one must be reserved for the desired signal, meaning that the remaining $K-1$ interfering signals must align and appear in the other direction. This process of packing the interfering signals into a smaller number of dimensions is the essence of interference alignment.  

The role of interference alignment can be better understood by considering the two extreme cases: $(M_{t},M_{r}) = (K,1)$ and $(M_{t},M_{r}) = (1,K)$. Recall that the  objective is to construct beamforming matrices satisfying the structural constraints and 
\[
\bU^{\top}\bH\bV = \bI.
\]
When $M_{t} = K$, then $\bV$ can be full matrix. Therefore, we can choose the beamforming matrices as $\bV = \bH^{-1}$ and $\bU = \bI$ corresponding to transmit zero-forcing. Similarly, if $M_{r} = K$, then we can choose the beamforming matrices as $\bV = \bI$ and $\bU = \bH^{-1}$ corresponding to receive zero-forcing. The concepts of transmit zero-forcing and receive zero-forcing are well understood in the communication theory literature. The reason why $M_{t} = K$ or $M_{r} = K$ works is the following. In both the cases, there are $K-1$ additional antennas at each transmitter or at each receiver to avoid interference. Essentially either the transmitters or the receivers take the burden to avoid interference. The condition $M_{t} + M_{r} \geq K+1$ says that this burden to avoid interference does not have to be taken solely either by the transmitters or the receivers, but can be shared by both. In other words, interference alignment can be thought of as a generalized zero-forcing strategy that allows the burden of interference avoidance to be shared by the transmitters and receivers by carefully designing the beams. The disadvantage of doing so is that, while the design of transmit or receive zero-forcing beams requires only local channel knowledge, the design of interference alignment beams requires global channel knowledge and even the computational aspects become more complicated. Since the existence proofs are nonconstructive, it is not clear if there is any closed-form algorithm or even iterative algorithm to numerically compute the interference alignment beams.

\subsection{Closed-Form Algorithm}
We showed that a linear beamforming strategy based on interference alignment achieves $K$ DoF whenever $M_t$ and $M_r$ satisfy $M_t + M_r \geq K + 1$. The proof of Theorem~\ref{thm:comp-fulldof} is not constructive. In this section, we consider the problem of numerical computation of interference alignment beams, i.e., computation of matrices $\bV$ and $\bU$ that satisfy the structural constraints imposed by transmit sets and receive sets, and diagonalize the channel matrix $\bH$
\be{eq:comp-v-u}
\bU^{\top}\bH\bV = \bI.
\ee
In the previous section, we have seen that the problem is easy if either $M_{t} = K$ or $M_{r} = K$, where the beamforming matrices correspond to either transmit zero-forcing or receive zero-forcing. In this section, we show that there exists a closed form solution when $M_{t} = K-1$ or $M_{r} = K-1$. Without any loss of generality, we consider the case $M_{t} = K-1$ and $M_{r} = 2$, and show that the closed-form solution described in Algorithm~\ref{algo:min2} satisfies the structural constraints and \eqref{eq:comp-v-u}. The rest of this section focuses on justifying the steps in Algorithm~\ref{algo:min2}.

\begin{algorithm}[h]
\caption{Closed Form Solution: $M_t = K-1$ and $M_r = 2$}
\label{algo:min2}
\begin{algorithmic}
\item[1:] For each $k \in \sK$, define the alignment matrix
\[
\bB_{k} = \bH(\sT_{k+2},\sT_{k+1})^{-1}\bH(\sT_{k+2},\sT_{k})
\]
\item[2:] Choose $\bv_{1}$ as an eigenvector of the matrix 
\[
\bB_{K}\bB_{K-1}\cdots\bB_{1}
\]
\item[3:] For $k = 1,2,\cdots,K-1$, compute
\[
\bv_{k+1} = \bB_{k}\bv_{k}
\]
\item[5:] Compute the transmit beamforming matrix $\bV$ such that 
\[
\bv_{k} = \bV(\sT_{k},k), \, \forall k \in \sK.
\]

\item[6:] Compute the receive beamfoming matrix $\bU = (\bH\bV)^{-\top}$.
\end{algorithmic}
\end{algorithm}
%

The usual approach to solve for $\bU$ and $\bV$ is by first eliminating $\bU$ by obtaining the necessary and sufficient conditions on $\bV$ for an appropriate $\bU$ to exist, and then solving for $\bV$. Let $\bM$ denote the matrix $\bH\bV$. 
We now obtain the necessary and sufficient conditions on the matrix $\bM$ so that its inverse $\bM^{-1} = \bU^{\top}$ satisfies the structural constraints imposed by the receive sets. For example, if $M_{r} = 2$, then the receive beamforming matrix should have the following structure:
\be{eq:comp-u-structure}
\bU = \mat{
\times &  &  &  & \times \\
\times & \times &  &  &  \\
 & \times &  \times &  &  \\
 & & \ddots & \ddots \\
 & & & \times & \times
}.
\ee
The nullity theorem \cite{fielder86,gilbert04} from linear algebra is useful in obtaining the neccesary and sufficient conditions on $\bM$.

\begin{lemma}[Nullity Theorem]
Complementary submatrices of a matrix and its inverse have the same nullity.
\end{lemma}

Two submatrices are {\em complementary} when the row numbers not used in one are the column numbers used in the other. For any subsets $\sA, \sB \subseteq \sK$, applying the Nullity Theorem to $\bM$ and $\bU^{\top} = \bM^{-1}$, we have that
\[
\begin{split}
\mathrm{nullity}\bM(\sA,\sB) & = \mathrm{nullity}\bU^{\top}(\sB^{c},\sA^{c}) \\
\Leftrightarrow |\sB| - \mathrm{rank}\bM(\sA,\sB) & = |\sA^{c}| - \mathrm{rank}\bU(\sA^{c},\sB^{c}) \\
\Leftrightarrow \mathrm{rank}\bM(\sA,\sB) & = \mathrm{rank}\bU(\sA^{c},\sB^{c}) + |\sA| + |\sB| - K.
\end{split}
\]
Observe that the structural constraints on the matrix $\bU$ can be described as
\be{eq:comp-u-descr-col}
\mathrm{rank}\bU(\sR_{k}^{c},k) = 0, \, \forall k \in \sK.
\ee
By choosing $\sA = \sR_{k}$ and $\sB = \co{k}$, we observe that structural constraints on $\bU$ are equivalent to the following constraints on $\bM$:
\be{eq:comp-alignment-conds}
\mathrm{rank}\bM(\sR_{k},\co{k}) = M_{r}-1, \, \forall k \in \sK.
\ee
Note that the above conditions are nothing but the interference alignment conditions. The matrix $\bM = \bH\bV$ should be interpreted as the matrix containing the receive directions as the columns
\be{eq:comp-fulldof-M}
\bM = \mat{\bH(\sK,\sT_{1})\bv_{1} & \bH(\sK,\sT_{2})\bv_{2} & \cdots & \bH(\sK,\sT_{K})\bv_{K}}
\ee
where $\bv_{k} \in \mathbb{C}^{M_{t}\times 1}$ denotes the beamforming vector corresponding to the message $W_{k}$, i.e., $\bv_{k} = \bV(\sT_{k},k)$. Consider the decoder of message $W_{k}$ which has access to the signals received by the receivers $\sR_{k}$. The submatrix 
\[
\bM(\sR_{k},\sK) = \mat{\bH(\sR_{k},\sT_{1})\bv_{1} & \bH(\sR_{k},\sT_{2})\bv_{2} & \cdots & \bH(\sR_{k},\sT_{K})\bv_{K}}
\]
represents the matrix with the column denoting the directions along which the signals appear at the decoder $k$. Thus, we see that the condition \eqref{eq:comp-alignment-conds} is equivalent to saying that the interfering signals should occupy only $M_{r} - 1$ dimensions out of the available $M_{r}$ dimensions at decoder $k$, leaving one dimension for the signal. With this intuition, we could have arrived at the alignment conditions \eqref{eq:comp-alignment-conds} directly without invoking the nullity theorem.
However, the constraints \eqref{eq:comp-alignment-conds} do not directly lead to a closed-form solution. 

We now demonstrate the usefulness of the nullity theorem by deriving another set of equivalent conditions on $\bM$ that immediately lead to the closed-form solution described in Algorithm~\ref{algo:min2}. The crucial observation is the following. In the description \eqref{eq:comp-u-descr-col}, we noticed that each column of $\bU$ has $K - M_{r}$ zeros. Alternatively, we can use the fact that each row of $\bU$ has $K - M_{r}$ zeros to arrive at an alternate description of the structural constraints on $\bU$:
\[
\mathrm{rank}\bU( k-1, k \uparrow (K-M_{r})) = 0, \, \forall k \in \sK.
\]
By choosing $\sA = \co{k-1}$ and $\sB = \{k\uparrow K-M_{r}\}^{c} = (k-1)\downarrow M_{r}$, we observe that the structural constraints on $\bU$ are equivalent to following constraints on $\bM$:
\[
\mathrm{rank}\bM(\co{k-1},(k-1)\downarrow M_{r}) = M_{r}-1, \, \forall k \in \sK.
\]
For the special case of $M_{t} = K-1$ and $M_{r} = 2$, we have that $\sT_{k} = \co{k-1}$ and $(k-1)\downarrow M_{r} = \{k-1,k-2\}$. Using the expression \eqref{eq:comp-fulldof-M} for $\bM$, we see that the above conditions can be written as
\[
\mathrm{rank}\mat{\bH(\sT_{k},\sT_{k-1})\bv_{k-1} & \bH(\sT_{k},\sT_{k-2})\bv_{k-2}} = 1, \, \forall k \in \sK.
\]
For a generic $\bH$, the submatrix $\bH(\sT_{k},\sT_{k-1})$ is invertible, and hence the above conditions can equivalently be expressed as
\[
\bv_{k-1} \propto \bB_{k-2}\bv_{k-2}
\]
where $\bB_{k-2} = \bH(\sT_{k},\sT_{k-1})^{-1}\bH(\sT_{k},\sT_{k-2})$. Therefore, the transmit beams must be designed to satisfy
\[
\begin{split}
\bv_{2} & \propto \bB_{1}\bv_{1} \\
\bv_{3} & \propto \bB_{2}\bv_{2} \\
& \ \ \vdots \\
\bv_{K} & \propto \bB_{K-1}\bv_{K-1} \\
\bv_{1} & \propto \bB_{K}\bv_{K}.
\end{split}
\]
The above conditions are satisfied if and only if $\bv_{1}$ is an eigenvector of the matrix $\bB_{K}\bB_{K-1}\cdots\bB_{1}$, and $\bv_{k+1} \in \bB_{k}\bv_{k}$ for $k = 2,3,\cdots,K$. We can then compute the receive beamforming vectors by computing $\bM = \bH\bV$ and setting $\bU = \bM^{-\top}$. The choice of transmit beams and the nullity theorem ensures that the resulting receive beamforming matrix $\bU$ has the required structure \eqref{eq:comp-u-structure}.  
\subsection{Numerical Results}

\begin{figure}[th!]
\centering
\includegraphics[width=0.8\textwidth]{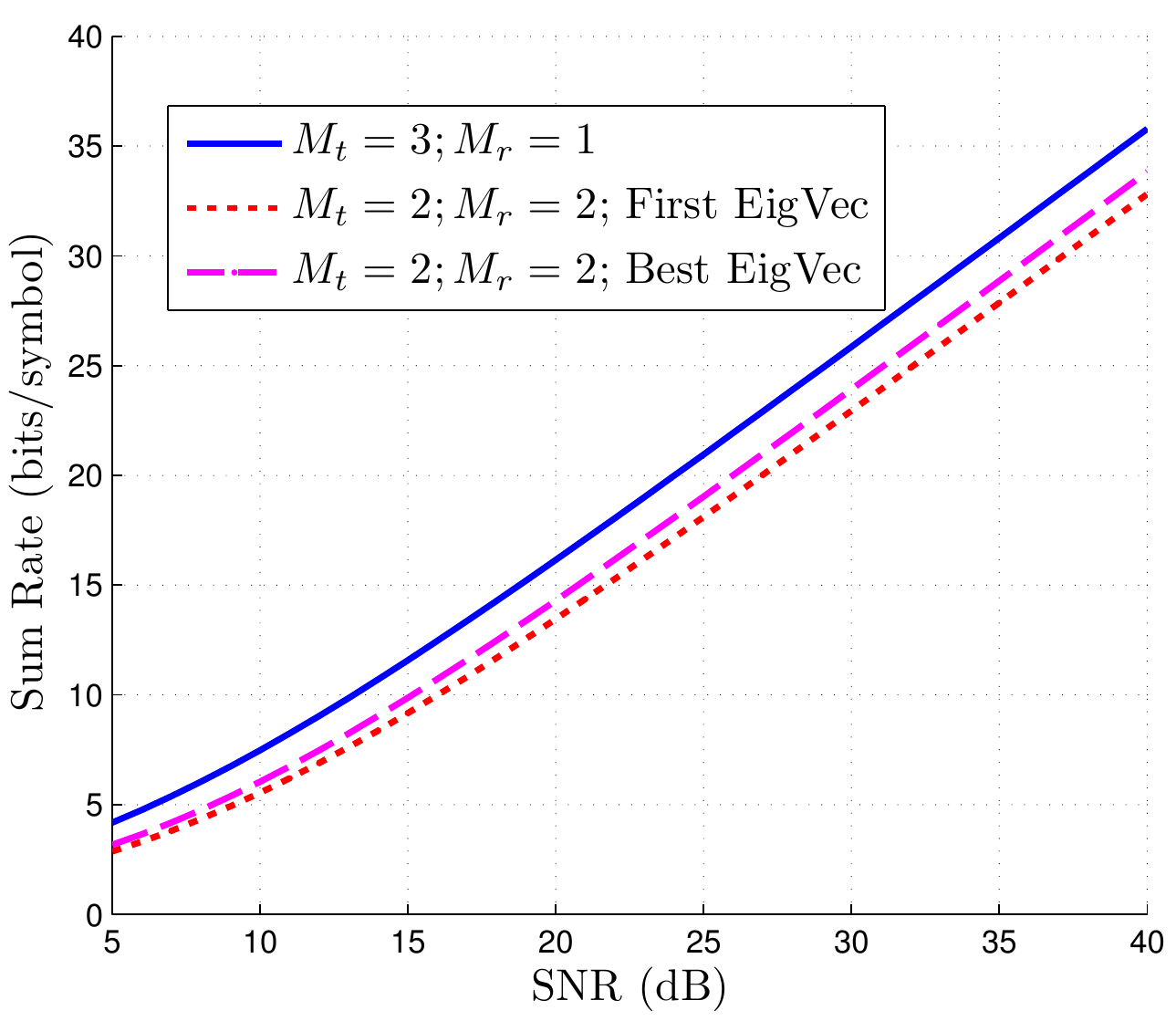}
\caption{Achievable sum-rates in a three-antenna system with alignment schemes.}
\label{fig:k3}
\end{figure}

In this section, we consider the three-antenna system, i.e., $K = 3$. From Theorem~\ref{thm:comp-fulldof}, we have that the maximum $3$ DOF is achievable if and only if $M_{t} + M_{r} \geq 4$. We numerically verify the achievability part of the theorem by showing that $3$ DoF is achievable when $M_{t} + M_{r} \geq 4$. Without any loss of generality, we only consider the two settings $(M_{t},M_{r}) = (3,1)$ and $(M_{t},M_{r}) = (2,2)$ because the other settings can be shown to follow from these two settings. In Figure~\ref{fig:k3}, we plot the average achievable sum-rate, where the averaging is performed over the multiple realizations of the channel coefficients which are generated independently according to complex normal distribution. When $(M_{t},M_{r}) = (3,1)$, the system is equivalent to a broadcast channel, and so we use the zero forcing transmit beams described in Section~\ref{sec:comp-fulldof-mimo}. When $(M_{t},M_{r}) = (2,2)$, we have that $M_{r} = K-1$, and so we use the alignment scheme described in Algorithm~\ref{algo:min2} to compute the transmit and receive beams. In step 2 of Algorithm~\ref{algo:min2}, the computation of the transmit beam $\bv_{1}$ involves computing an eigenvector of the $2 \times 2$ matrix. In Figure~\ref{fig:k3}, we plot the two curves for the setting $(M_{t},M_{r}) = (2,2)$: one corresponds to arbitrary eigenvector and the other corresponds to best eigenvector over each channel realization. 

The plots numerically verifiy that the achievable scheme described in Algorithm~\ref{algo:min2} indeed achieves $3$ DoF with $(M_{t},M_{r}) = (2,2)$. Indeed, a linear growth of $10$ bits/symbol in sum-rate for every $10$ dB improvement in SNR corresponds to
\[
\frac{10}{\log_{2}10} \approx 3 ~ \dof.
\]
It is also interesting to see that $(M_{t},M_{r}) = (3,1)$ achieves better sum-rate when compared to $(M_{t},M_{r}) = (2,2)$. The performance gap is roughly $3$ dB at high SNRs when arbitrary eigenvector is used, and is roughly $2$ dB when best eigenvector is used. 
\section{DoF with CoMP Transmission} \label{sec:comp-tx}
In the previous sections, we derived an outer bound on the DoF and showed that the DoF is equal to the maximum value $K$ if and only if $M_{t} + M_{r} \geq K+1$. In this section, we set $M_{r} = 1$, and consider the problem of characterizing $\dof(K,M_{t},1)$, the DoF of interference channel with CoMP transmission, as a function of $K$ and $M_{t}$. From the outer bound in Section~\ref{sec:comp-ob-sumdof}, we obtain that $\dof(K,M_{t},1)$ is upper bounded as
\[
\dof(K,M_{t},1) \leq 
\begin{dcases}
\frac{K+M_{t}-1}{2}, & K+M_{t} \textrm{ is odd} \\
\frac{K}{K-1}\frac{K+M_{t}-2}{2} \leq \left\lceil\frac{K+M_{t}-1}{2}\right\rceil, & K+M_{t} \textrm{ is even}.
\end{dcases}
\]
For the achievability part, we prove the following two theorems. For any $K$ and $M_{t}$, we propose a scheme that aims at achieving a DoF of $(K+M_{t}-1)/2$. A crucial part of the proof involves checking that a certain Jacobian matrix has full row rank. 
We could verify in MATLAB that the Jacobian matrix has full row rank for all the values of $K$ and $M_{t}$ that we checked. Specifically, we checked till $K \leq 9$, but we conjecture that the result holds true for any $K$ and $M_{t}$. For more discussion on the problematic issue, we refer the reader to Section~\ref{sec:comp-tx-mk-ai}.
\begin{theorem} \label{thm:comp-tx}
The DoF of interference channel with CoMP transmission satisfies
\[
\dof(K,M_{t},1) \geq \frac{K+M_{t}-1}{2}
\]
for all $M_{t} \leq K < 10$.
\end{theorem}
Combining the above theorem with the outer bound, we have determined the DoF exactly when $K + M_{t}$ is odd, and approximately when $K+M_{t}$ is even (for all $M_{t} \leq K < 10$). For the special case of $M_{t} = K-2$, we propose an achievable scheme that exactly meets the outer bound. 
\begin{theorem} \label{thm:comp-tx-mk2}
The DoF of interference channel with CoMP transmission with $M_{t} = K-2$ satisfies
\[
\dof(K,K-2,1) = \frac{KM_{t}}{M_{t}+1} = \frac{K(K-2)}{K-1}.
\]
\end{theorem}
Theorem~\ref{thm:comp-tx-mk2} is first published in \cite{annapureddy2010degrees} for the special case of $K = 4$ and $M_{t} = 2$ and in \cite{naguib2011achievable} for the general case. The proofs offered in both of the above papers are not complete. The central issue is in proving that a certain random matrix has full rank for generic channel coefficients. In this paper, we overcome this issue by exploiting the notion of algebraic independence. Before proving the above theorems, we first explain the connection  to the DoF of the MISO interference channel. 
\subsection{Relation to MISO Interference Channel} \label{sec:comp-tx-miso}

The MISO interference channel with $N_{t} = M_{t}$ antennas per transmitter and the cellular uplink channel with $M_{t}$ number of users per cell are similar to the interference channel with CoMP transmission in the sense that, in all the three channels, each message is transmitted using $M_t$ antennas and received using only one antenna. The difference is that the messages share the antennas in the CoMP channel, whereas the messages have dedicated antennas in the other two channels. Both the MISO interference channel and the cellular uplink channel have the same DoF, equal to $KM_{t}/(M_{t}+1)$ for all $M_{t} < K$. In comparison, we see that the interference channel with CoMP transmission has a smaller DoF except in the special cases where $M_{t} \in \{1,K-1,K-2,K\}$.
\begin{claim} \label{claim:comp-tx-miso}
For all $M_{t} \not\in \{1,K-2,K-1,K\}$, 
\[
\dof(K,M_{t},1) < \frac{KM_{t}}{M_{t}+1}.
\]
\end{claim}
\begin{IEEEproof}
Suppose $M_{t} + K$ is odd. Then we see that
\[
\begin{split}
\dof(K,M_{t},1) = \frac{K+M_{t}-1}{2} & < \frac{KM_{t}}{M_{t}+1} \\
\Leftrightarrow K(M_{t}+1)+(M_{t}-1)(M_{t}+1) & < 2KM_{t} \\
\Leftrightarrow (M_{t}-1)(M_{t}+1) & < K(M_{t}-1) \\
\Leftrightarrow M_{t} + 1 & < K
\end{split}
\]
which is true since we assumed that $M_{t} < K-2$. Suppose $M_{t} + K$ is even; then
\[
\begin{split}
\dof(K,M_{t},1) \leq \frac{K}{K-1}\frac{K+M_{t}-2}{2} & < \frac{KM_{t}}{M_{t}+1} \\
\Leftrightarrow K(M_{t}+1)+(M_{t}-2)(M_{t}+1) & < 2KM_{t}-2M_{t} \\
\Leftrightarrow (M_{t}-1)(M_{t}+2) & < K(M_{t}-1) \\
\Leftrightarrow M_{t}+2 & < K
\end{split}
\]
which is true since we assumed that $M_{t} < K-2$.
\end{IEEEproof}
We now proceed to prove Theorems~\ref{thm:comp-tx} and \ref{thm:comp-tx-mk2}.
 \section{CoMP Transmission: Proof of Theorem~\ref{thm:comp-tx-mk2}} \label{sec:comp-tx-mk2} 
 
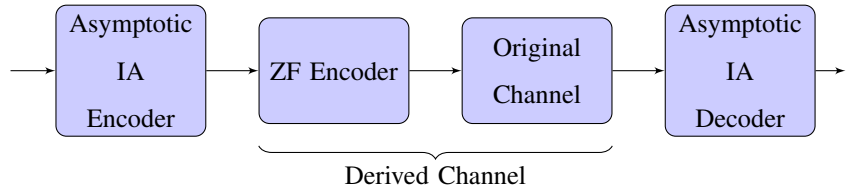
\begin{figure*}[th!] 
\centering 
\tikzstyle{block} = [rectangle, draw, fill=blue!20,  
    text width=5em, text centered, rounded corners, minimum height=4em] 
\tikzstyle{line} = [draw, -latex'] 
\begin{tikzpicture}[node distance = 2.7cm, auto] 
            \node [block] (H) {Original Channel}; 
           \node [block,left of=H] (ZFE) {ZF Encoder}; 
            \node [block,left of=ZFE] (IAE){ Asymptotic IA \\ Encoder}; 
        \node [block,right of=H] (IAD){Asymptotic IA \\ Decoder}; 
        \path[line] (-7,0) -- (IAE); 
            \path[line] (IAE) -- (ZFE); 
        \path[line] (ZFE) -- (H); 
        \path[line] (H) -- (IAD); 
        \path[line] (IAD) -- (4.1,0); 
        \draw [black,decorate,decoration={brace,amplitude=5pt}] 
                        (1,-1)  -- (-3.7,-1)  node [black,midway,below=4pt] {Derived Channel}; 
\end{tikzpicture} 
\caption{Summary of the achievable scheme.} 
\label{fig:ZF-IA} 
\end{figure*}

In this section, we show that the DoF of the interference channel with CoMP transmission and a transmit cooperation order of $M_{t} = K-2$ and a receive cooperation order $M_{r}=1$ is equal to 
\[
\dof(K,K-2,1) = \frac{KM_{t}}{M_{t}+1}. 
\]
The achievable scheme is based on transmit and receive beamforming. As summarized in Figure~\ref{fig:ZF-IA}, the beam design process is broken into two steps. First, we transform each parallel CoMP channel into a derived channel. Then, we design an asymptotic interference alignment scheme over the derived channel achieving the requied DoF in an asymptotic fashion with the number of parallel channels $L \rightarrow \infty$.

\subsection{Derived Channel}

Recall from Section~\ref{sec:comp-tx-miso}, that the cellular uplink channel with $M_{t}$ transmitters per cell has $KM_{t}/(M_{t}+1)$ DoF. Therefore, we first transform the CoMP channel into a derived channel that mimics the cellular uplink channel. For each $k$, the transmit set $\sT_k = k \uparrow M_{t}$  of user $k$ consists of $M_{t}$ transmitters. We use the $M_{t}$ transmitters in $\sT_k$ to create $M_{t}$ virtual transmit nodes with inputs $X_k^{(1)}, X_{k}^{(2)},\cdots,X_{k}^{(M_{t})}$. The channel inputs of the CoMP channel are related to the channel inputs of the derived channel through a linear transformation. The contribution of the derived channel inputs $X_k^{(1)}, X_{k}^{(2)},\cdots,X_{k}^{(M_{t})}$ in the real transmit signals $X_{k},X_{k+1},\cdots,X_{k+M_{t}-1}$ is defined by a $M_{t} \times M_{t}$ beamforming matrix; i.e.,
\[
\mat{X_{k} \\ X_{k+1} \\ \vdots \\ X_{k+M_{t}-1}} = (*) +  \bV_{k} \mat{X_{k}^{(1)} \\ X_{k}^{(2)} \\ \vdots \\ X_{k}^{(M_{t})}}
\]
where $*$ represents the contribution from the derived channel inputs of other users. Thus, we see that the beamforming matrices $\bV_{1},\bV_{2},\cdots,\bV_{K}$, which will be specified later, define the transformation from the original channel to the derived channel. The message $W_k$ of user $k$ is divided into $M_{t}$ parts 
\[
W_{k} = \lt(W_k^{(1)}, W_{k}^{(2)},\cdots,W_{k}^{(M_{t})}\rt)
\]
such that the $m$th part controls the derived channel input $X_k^{(m)}$. Thus we can treat the virtual transmit nodes as non-cooperative transmitters communicating to the same receiver and so this system is similar to a cellular uplink system with $M_{t}$ trasmitters per cell:
\be{eq:comp-dc} 
Y_{i} = \sum_{k=1}^{K}\sum_{m=1}^{M_{t}}g_{ik}^{(m)}X_{k}^{(m)}+Z_{i}, \, i \in \sK
\ee 
where $g_{ik}^{(m)}$ represents the derived channel coefficient from transmitter $m$ in cell $k$ to the receiver in cell $i$. It is easy to see that the derived channel coefficients are related to the original channel coefficients as
\[
\mat{g_{ik}^{(1)} & g_{ik}^{(2)} & \cdots & g_{ik}^{(M_{t})}} = \bH(i,\sT_{k})\bV_{k}
\]
for all $i,k \in \sK$, where $\bH$ denotes $K \times K$ channel transfer matrix of the CoMP channel.

\subsection{Generic Channel Coefficients}
The derived channel \eqref{eq:comp-dc} is similar to the cellular uplink channel with $K$ cells and $M_t$ transmitters in each cell, which has $KM_t/(M_t+1)$ DoF with generic channel coefficients \cite{jafar2010asymptotic}. A naive argument is to conclude from here that the derived channel, and hence the CoMP channel with generic channel coefficients, also has the same DoF. However, from Claim~\ref{claim:comp-tx-miso} in Section~\ref{sec:comp-tx-miso}, we know that the DoF of the CoMP channel is strictly smaller than $KM_t/(M_t+1)$, which means that the above naive argument has to be incorrect. 

The reason for the failure of the above naive argument is related to the subtle concept of generic channel coefficients. Indeed, the derived channel has $KM_t/(M_t+1)$ DoF with generic channel coefficients, which means that there exists a nonzero polynomial $f_{g}(\bg)$ in the derived channel coefficients
\[
\bg = \{g_{ij}^{(m)}(l): 1 \leq i,j \leq K, 1 \leq m \leq M_{t}, 1 \leq \ell \leq L \}
\]
such that the achievable scheme works for all $\bg$ such that $f_{g}(\bg) \neq 0$. In the case of the cellular uplink channel, this statement makes sense since the coefficients $\bg$ are generated by nature and hence can be assumed to be generic. However, in the case of the CoMP channel, nature generates the original channel coefficients $\{h_{ij}(l)\}$, denoted by $\bh$. The coefficients $\bg$ are derived from $\bh$ using rational transformations. Suppose we expand the polynomial $f_g$ in terms of the coefficients $\bh$ to obtain the rational function $f_h(\bh) = f_g(\bg(\bh))$. There are two possibilities: the function $f_h$ is either identically equal to zero or it is nonzero. If $f_{h} = 0$, then the achievable scheme designed for the derived channel with generic $\bg$ may fail for all realizations of $\bh$, in which case the DoF result of the derived channel with generic channel coefficients {\em cannot} be directly applied to CoMP channel with generic channel coefficients. On the other hand, if $f_{h}$ is a nonzero function, then we see that the achievable scheme works for generic $\bh$, in which case the DoF result of the derived channel with generic channel coefficients {\em can} be directly applied to CoMP channel with generic channel coefficients. 

In summary, we need to be careful in applying the DoF result of the cellular uplink channel to the CoMP channel, and the applicability of the result depends on how the derived channel coefficients are related to the original channel coefficients. 
\subsection{Zero-Forcing Step} \label{sec:comp-tx-zf}
\begin{figure}[th!] 
\centering 
\begin{tikzpicture} 
\tikzstyle{tx}=[rectangle,fill=black!30] 
\tikzstyle{rx}=[rectangle,fill=black!30] 
\def\K{4}\def\M{2}\def\Mm{1} \def\sep{6} 
\colorlet{dircolor}{green!50!black} 
\colorlet{intcolor}{red} 
 
        \foreach \name / \i in {1,...,\K}{ 
                       \node[tx] (t1-\name) at (0,-0.1-2*\i) {$X^{(1)}_\i$}; 
                 \node[tx] (t2-\name) at (0,-0.9-2*\i) {$X^{(2)}_\i$}; 
                \node[rx](r-\name) at (\sep,-0.5-2*\i) {$Y_\i$};} 
        \foreach \j in {1,...,\K}{ 
                   \path[dircolor] (t1-\j) edge (r-\j); 
                \path[dircolor] (t2-\j) edge (r-\j);} 
                 
                \path[dashed,intcolor] (t1-2) edge (r-1); 
                \path[dashed,intcolor] (t2-2) edge (r-1); 
                \path[dashed,intcolor] (t1-3) edge (r-2); 
                \path[dashed,intcolor] (t2-3) edge (r-2); 
                \path[dashed,intcolor] (t1-4) edge (r-3); 
                \path[dashed,intcolor] (t2-4) edge (r-3); 
                \path[dashed,intcolor] (t1-1) edge (r-\K); 
                \path[dashed,intcolor] (t2-1) edge (r-\K); 
                 
                \path[dotted,intcolor] (t1-1) edge (r-2); 
                \path[dotted,intcolor] (t2-1) edge (r-3); 
                \path[dotted,intcolor] (t1-2) edge (r-3); 
                \path[dotted,intcolor] (t2-2) edge (r-4); 
                \path[dotted,intcolor] (t1-3) edge (r-4); 
                \path[dotted,intcolor] (t2-3) edge (r-1); 
                \path[dotted,intcolor] (t1-4) edge (r-1); 
                \path[dotted,intcolor] (t2-4) edge (r-2); 
                         
\end{tikzpicture} 
\caption{The derived channel in Section~\ref{sec:comp-tx-zf} when $K = 4$ and $M_t = 2$. The thick green lines indicate the links carrying signal. The dashed and dotted red lines indicate the links carrying interference. Dotted lines indicate that the corresponding coefficients are equal to $1$.} 
\label{fig:k4m2} 
\end{figure}
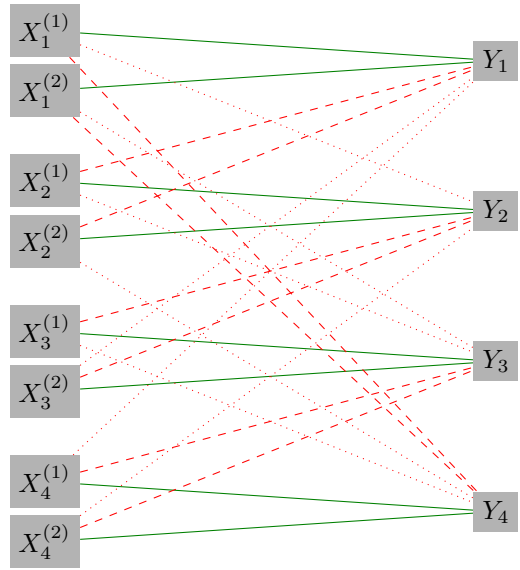

We now specify our choice of the beamforming matrices $\bV_{1},\bV_{2},\cdots,\bV_{K}$, that define the relation of the derived channel coefficients to the original channel coefficients. As we shall notice later during the design of the asymptotic interference alignment scheme, the beamforming matrices should be chosen to minimize the number of nontrivial derived channel coefficients, where we say that a derived channel coefficient is trivial if it is equal to either zero or one. Therefore, the objective is to set as many derived channel coefficients as possible to zeros or ones. Consider the derived channel coefficients
\[
\mat{
g_{k+1,k}^{(1)} & g_{k+1,k}^{(2)} & \cdots & g_{k+1,k}^{(M_{t})} \\
g_{k+2,k}^{(1)} & g_{k+2,k}^{(2)} & \cdots & g_{k+2,k}^{(M_{t})} \\
\vdots & \vdots & \ddots & \vdots \\
g_{k+M_{t},k}^{(1)} & g_{k+M_{t},k}^{(2)} & \cdots & g_{k+M_{t},k}^{(M_{t})} } = \bH(\sT_{k+1},\sT_{k})\bV_{k}.
\]
By choosing $\bV_{k} = \bH(\sT_{k+1},\sT_{k})^{-1}$, we can set all the above mentioned derived channel coefficients to either zero or one. In particular, we see that for each $i \in \sT_{k+1}$
\[
g_{ik}^{(m)} = 
\begin{dcases}
1 & i = k + m \\
0 & \text{Otherwise}.
\end{dcases}
\]
Since we assumed that $M_{t} = K-2$, the set $\sT_{k+1}$ contains all the receiver indices except for $k-1$ and $k$. Therefore, we see that each transmitter $X_{k}^{(m)}$ in the derived channel causes interference to only two receivers, i.e., receivers $k+m$ and $k-1$. Thus, the derived channel \eqref{eq:comp-dc} can be simplified as 
\be{eq:comp-dc2} 
Y_{i} = \sum_{m = 1}^{M_{t}}g_{ii}^{(m)}X_{i}^{(m)} +\sum_{m = 1}^{M_{t}}g_{i,i+1}^{(m)}X_{i+1}^{(m)}+ \sum_{m =1}^{M_{t}} X_{i - m}^{(m)} + Z_{i} 
\ee 
where the coefficients $g_{ii}^{(m)}$ and $g_{i,i+1}^{(m)}$ are given by 
\be{eq:comp-h2g} 
\begin{split} 
\mat{g_{i,i+1}^{(1)} & \cdots & g_{i,i+1}^{(M_{t})}} & = \bH(i,\sT_{i+1})\bV_{i+1} = \bH(i,\sT_{i+1})\bH(\sT_{i+2},\sT_{i+1})^{-1} \\ 
\mat{g_{ii}^{(1)} & \cdots & g_{ii}^{(M_{t})}} & = \bH(i,\sT_{i})\bV_{i} = \bH(i,\sT_{i})\bH(\sT_{i+1},\sT_{i})^{-1}. 
\end{split} 
\ee 
Figure~\ref{fig:k4m2} provides a description of the derived channel for the special case of $K = 4$ and $M_t = 2$.

\subsection{Asymptotic Interference Alignment} \label{sec:comp-tx-mk2-ia}

In this section, we consider $L$ parallel derived channels and propose a scheme achieving a DoF that is arbitrary close to $KM_{t}/(M_{t}+1)$ in the limit $L \rightarrow \infty$. We can combine the $L$ parallel channels of \eqref{eq:comp-dc2} and express them together as
\be{eq:comp-dc-sym} 
\uy_{i} = \sum_{m = 1}^{M_{t}}\bG_{ii}^{(m)}\ux_{i}^{(m)} +\sum_{m = 1}^{M_{t}}\bG_{i,i+1}^{(m)}\ux_{i+1}^{(m)}+ \sum_{m =1}^{M_{t}} \ux_{i - m}^{(m)} + \uz_{i} 
\ee 
where $\ux^{(m)}_{j}, \uy_{i}$ and $\uz_{i}$ are $L \times 1$ column vectors and $\bG_{ij}^{(m)}$ is $L \times L$ diagonal channel transfer matrix given by
\[
\bG_{ij}^{(m)} = \mat{g_{ij}^{(m)}(1) \\ & g_{ij}^{(m)}(2) \\ & & \ddots \\ & & & g_{ij}^{(m)}(L)}.
\] 
The achievable scheme that we propose is based on the asymptotic alignment  scheme introduced by Cadambe and Jafar in \cite{cadambe2008interference}.

\begin{definition}[Cadambe-Jafar (CJ) subspace] 
The order-$n$ CJ subspace generated by the diagonal matrices $$\bG_{1},\bG_{2},\cdots,\bG_{N}$$ is defined as the linear subspace spanned by the vectors 
\[ 
\{\bG_{1}^{a_{1}}\bG_{2}^{a_{2}}\cdots\bG_{N}^{a_{N}}\mathbf{1}: \ba \in \mathbb{Z}_{+}^{N} \text{ and } \sum_{i} a_{i} \leq n\}. 
\] 
The matrix containing these ${N+n \choose n}$ vectors as columns is said to be the order-$n$ CJ matrix. 
\end{definition} 
 
Let $\bV$ denote the order-$n$ CJ supspace (and the corresponding matrix) generated by the nontrivial channel matrices carrying interference:
\[
\{\bG_{i,i+1}^{(m)}: i \in \sK, 1 \leq m \leq M_{t}\}. 
\] 
We use $\bV$ as the transmit beamforming matrix at every transmitter. The nice property about the CJ subspace is that the interference seen at any receiver is limited to the order-$(n+1)$ CJ subspace, denoted by $\mathbf{INT}$. At receiver $k$, the desired signal streams appear along the directions 
\[ 
\mat{\bG_{kk}^{(1)}\bV & \bG_{kk}^{(2)}\bV & \cdots & \bG_{kk}^{(M_{t})}\bV}. 
\] 
The proposed scheme works if the receivers are able to extract out the desired signal streams free of interference, which is true if the matrix 
\be{eq:mat-mk-2} 
\bM_{k} = \mat{\bG_{kk}^{(1)}\bV & \bG_{kk}^{(2)}\bV & \cdots & \bG_{kk}^{(M_{t})}\bV & \int} 
\ee 
has full column rank for every $k \in \sK$. For the matrix $\bM_{k}$ to have full column rank, the number of rows, equal to the number of parallel channels $(L)$, must be greater than or equal to the number of columns. The number of columns in $\bV$ and $\int$, respectively, is given by 
\[ 
\begin{split} 
|\bV| = & \ {KM_{t} + n \choose KM_{t}} \\ 
|\int| = & \ {KM_{t}+n+1 \choose KM_{t}}. 
\end{split} 
\] 
Hence the number of columns in $\bM_{k}$ is equal to $M_{t}|\bV| + |\int|$ . We set  $L = M_{t}|\bV| + |\int|$ so that $\bM_{k}$ is a square matrix for each $k \in \sK$. Note that the matrix $\bM_k$ depends on the derived channel coefficients 
\[
g_{ii}^{(m)}(\ell), g_{i,i+1}^{(m)}(\ell): 1 \leq m \leq M_t, 1 \leq i \leq K, 1 \leq  \ell \leq L.
\] 
We need to prove that the matrices $\bM_1, \cdots,\bM_k$ have full rank for generic (original) channel coefficients
\[
h_{ij}(\ell): 1 \leq i,j \leq K, 1 \leq  \ell \leq L.
\] 
The proof uses techniques from algebraic geometry summarized in Section~\ref{sec:app-ag}. Using Corollary~\ref{corr:app-ag-fullrank}, we see that the matrices $\bM_{1},\bM_{2},\cdots,\bM_{k}$ have full column rank if the rational transformation \eqref{eq:comp-h2g} from the original channel coefficients to the derived channel coefficients is such that the rational functions denoted by the variables 
\be{eq:gij-rxk} 
g_{kk}^{(m)}, g_{i,i+1}^{(m)}: 1 \leq m \leq M_t, 1 \leq i \leq K
\ee 
are algebraically independent. Before we prove the algebraic independence, we show that the proposed scheme achieves the required DoF. Since the derived channel has a total of $KM_{t}$ number of transmitters, and the proposed interference alignment scheme creates $|\bV|$ number of interference-free AWGN channels per each transmitter, we obtain the following lower bound on the (normalized) DoF:
\[
\begin{split}
\dof(K,K-2,1,L) \geq \frac{KM_{t}|\bV|}{L} = \frac{KM_{t}|\bV|}{M_{t}|\bV| + |\int|} = \frac{KM_{t}}{M_{t} + 1 + \frac{KM_{t}}{n+1}}.
\end{split}
\]
Therefore, we obtain that
\[
\begin{split}
\dof(K,K-2,1) & = \limsup_{L \rightarrow \infty}\dof(K,K-2,1,L) \\
&  \geq \lim_{n \rightarrow \infty} \frac{KM_{t}}{M_{t} + 1 + \frac{KM_{t}}{n+1}} \\
& = \frac{KM_{t}}{M_{t}+1}.
\end{split}
\]
\subsection{Proof of Algebraic Independence}  \label{sec:comp-tx-mk2-ag}
Since the achievable scheme is symmetric across the user indices, it is sufficient to prove the claim for $k = 1$. The $(K+1)M_t$ variables \eqref{eq:gij-rxk} are rational functions of the $K^2$ variables $\{h_{ij}: 1 \leq i,j \leq K\}$.  Let $\bJ$ denote the corresponding $(K+1)M_t \times K^2$ Jacobian matrix. From Lemma~\ref{lemma:app-ag-jacob}, the variables \eqref{eq:gij-rxk} are algebraically independent if and only if the Jacobian matrix $\bJ$ has full row rank equal to $(K+1)M_t$. Let 
\be{eq:subJ} 
\bJ[\bg_0,\bg_1,\cdots,\bg_K;\bh_0,\bh_1,\cdots,\bh_K] 
\ee 
denote the $(K+1)M_{t} \times (K+1)M_{t}$ submatrix of $\bJ$ with rows corresponding to the variables $\bg_0,\bg_1,\cdots,\bg_K$ and columns corresponding to the variables $\bh_0,\bh_1,\cdots,\bh_K$, where 
\[
\begin{split} 
& \bg_0 = \lt(g_{11}^{(1)},g_{11}^{(2)}, \cdots, g_{11}^{(M_t)} \rt) \\  
& \bg_i = \lt(g_{i,i+1}^{(1)},g_{i,i+1}^{(2)}, \cdots, g_{i,i+1}^{(M_t)} \rt) \\ 
& \bh_0 = \lt(h_{11}, h_{22}, \cdots, h_{M_tM_t}\rt) \\  
& \bh_i = \lt(h_{i,i+1}, h_{i,i+2}, \cdots, h_{i,K},h_{i,1}, \cdots,h_{i,i-2}\rt). 
\end{split} 
\]
We complete the claim by showing that square matrix \eqref{eq:subJ} has full rank. This is easy to verify using the symbolic toolbox of MATLAB for any fixed $K$. An analytical proof involves computing the submatrix \eqref{eq:subJ} at a specific point $\bH = \bA$, and showing that it has full rank. Although this is true at any randomly generated $\bA$, certain choices can simplify the proof. We choose $\bA$ to be the circulant matrix given by 
\[
a_{ij} = \left\{ 
\begin{array}{cc} 
1 & \text{if } j = i \text{ or } j = i-1 \\ 
0 & \text{otherwise} 
\end{array}. 
\right. 
\] 
For the special case of $K = 4$ and $M_t = 2$, the matrix $\bA$ is given by 
\[
\bA = \mat{ 
1 & 0 & 0 & 1 \\ 
1 & 1 & 0 & 0 \\ 
0 & 1 & 1 & 0 \\  
0 & 0 & 1 & 1 
}.
\] 
The following claim, whose proof is relegated to Appendix~\ref{sec:comp-claim-jacob}, completes the proof of Theorem~\ref{thm:comp-tx-mk2}.  
\begin{claim} \label{claim:comp-jacob}
The determinant of the submatrix \eqref{eq:subJ} evaluated at the point $\bH = \bA$ is equal to $\pm 1$. 
\end{claim} 
\subsection{Discussion}
We end the section by explaining why the proposed scheme does not extend for arbitrary $M_{t} < K-2$. Observe that a straightforward extension of the achievable scheme involves the same choice of ZF transmit beams in Section~\ref{sec:comp-tx-zf}. However, since $M_{t} < K-2$, each transmitter in the derived channel now causes interference at $K-M_{t}$ receivers, i.e., the transmitter $X_{k}^{(m)}$ causes interference at the receivers $k+m, k+M_{t}+1, k + M_{t}+2, \cdots, k+K-1$. Since the asymptotic interference alignment scheme requires that we use all the nontrivial channel matrices in generating the CJ subspace, we can verify that the achievable scheme works if the rational functions defined by the variables
\[
g_{kk}^{(m)}, g_{i,i+1}^{(m)},g_{i,i+2}^{(m)}, \cdots, g_{i,i+K-M_{t}-1}^{(m)}: 1 \leq m \leq M_t, 1 \leq i \leq K
\]
are algebraically independent for each $k \in \sK$. The total number of rational functions is given by
\[
(1 + (K - M_{t} - 1)K)M_{t}.
\]
If the above number were to be greater than $K^{2}$, then we can end this discussion since $m > n$ rational functions in $n$ variables cannot be algebraically independent. But that is not the case. For example, when $M_{t} = 2$ and $K = 5$, we have $22$ rational functions in $25$ variables. If these rational functions were to be algebraically independent, then the achievable scheme generalizes achieving a DoF of $KM_{t}/(M_{t}+1)$, but we know from the discussion in Section~\ref{sec:comp-tx-miso} that the DoF is strictly less than $KM_{t}/(M_{t}+1)$ for all $1 < M_{t} < K-2$. Therefore, it must be that these rational functions are algebraically dependent. 

\section{CoMP Transmission: Proof of Theorem~\ref{thm:comp-tx}} \label{sec:comp-tx-mk}

In this section, we show that the DoF of the interference channel with CoMP transmission and with transmit cooperation order of $M_{t}$ and a receive cooperation order of $M_{r} = 1$ is lower-bounded by
\[
\dof(K,M_t,1) \geq \frac{K+M_{t}-1}{2}. 
\]
We prove this by arguing that the DoF vector 
\[
d_i = \left\{\begin{array}{ll} 1 & 1 \leq i \leq M_t-1\\ 0.5 & M_t \leq i \leq K \end{array} \right. 
\]
is achievable; i.e., the first $M_t-1$ users benefit from cooperation and achieve $1$ degree of freedom, whereas the remaining $K - M_t + 1$ users achieve $1/2$ degree of freedom  just like in the interference channel without cooperation. Conceptually, the achievable scheme in this section is identical to the achievable scheme in Section~\ref{sec:comp-tx-mk2} for the special case when $M_t = K-2$; i.e., the achievable scheme is again based on converting the CoMP channel into a derived channel and then employing the asymptotic interference alignment scheme on the derived channel, as summarized in Figure~\ref{fig:ZF-IA}. 
\subsection{Derived Channel} 

As in Section~\ref{sec:comp-tx-mk2}, we convert the CoMP channel into a derived channel that mimics the cellular uplink channel. Since our objective is to achieve a DoF vector that is asymmetric, the derived channel is also chosen to be asymmetric. The derived channel we consider in this section has two transmitters in each of the first $M_{t}-1$ cells, and one transmitter in the remaining $K-M_{t}+1$ cells. 
\be{eq:dc-km} 
\begin{split} 
Y_{i} & = \sum_{j=1}^{K} g_{ij}^{(1)} X_{j}^{(1)} + \sum_{j=1}^{M_{t}-1} g_{ij}^{(2)}X_{j}^{(2)} + Z_{i}. 
\end{split} 
\ee 
As in Section~\ref{sec:comp-tx-mk2}, we assume that the channel inputs of the CoMP channel are related to the channel inputs of the derived channel through a linear transformation. The contribution of the derived channel input $X_{j}^{(m)}$ in the real transmit signals $X_{j},X_{j+1},\cdots,X_{j+M_{t}-1}$ is defined by a $M_{t} \times 1$ beamforming vector, i.e., 
\[
\mat{X_{j} \\ X_{j+1} \\ \vdots \\ X_{j+M_{t}-1}} = (*) +  \bv_{j}^{(m)} X_{j}^{(m)}
\]
where $*$ represents the contribution from other derived channel inputs. It is easy to see that the derived channel coefficients are related to the original channel coefficients as
\[
g_{ij}^{(m)}= \bH(i,\sT_{j})\bv_{j}^{(m)}
\]
for all $i,j \in \sK$ and appropriate $m$. Since we are designing the achievable scheme to achieve $1$ degree of freedom for the first $M_{t}-1$ users, it must be that the first $M_{t}-1$ receivers in the derived channel do not see any interference. 
\subsection{Zero-Frocing Step} \label{sec:zf-km} 
We now explain our choice of the beamforming vectors that ensures that the first $M_{t}-1$ receivers do not see any interference. 
\subsubsection{ZF beam design} \label{sec:zf-idea} 
We first describe the general idea of constructing a zero-forcing beam. Consider the problem of designing a zero-forcing beam $\bv$ to be transmitted by $n$ transmit antennas indexed by the set $\sT \subseteq \sK$ such that it does not cause interference at $n-1$ receive antennas indexed by the set  $\sI \subseteq \sK$, i.e., 
\[
\bH(\sI,\sT)\bv =\bo. 
\] 
Since $\bH(\sI,\sT)$ is a $n-1 \times n$ matrix, the choice for $\bv$ is unique up to a scaling factor. For any arbitrary row vector $\ba$ of length $n$, we can use the Laplace expansion to expand the determinant
\[
\det{\mat{\bH(\sI,\sT) \\ \ba}} = \sum_{j=1}^{n} a_{j} c_{j}
\]
where $c_{j}$ is the cofactor of $a_{j}$, that depends only on the channel coefficients in $\bH(\sI,\sT)$, and is independent of $\ba$. By setting the beamforming vector $\bv$ as $\bv = [c_{1} \ c_{2} \ \cdots \ c_{n}]$, we see that an arbitrary receiver $i$ sees the signal transmitted along the beam $\bv$ with a strength equal to 
\[
\bg = \bH(i,\sT)\bv = \det{\mat{\bH(\sI,\sT) \\ \bH(i,\sT)}} = \det{\bH(\sI \cup i, \sT)}.
\]
Clearly, this satisfies the zero-forcing condition $\bH(i,\sT)\bv = 0$ for all $i \in \sI$.
\subsubsection{Design of transmit beam $\bv_{j}^{(1)}$ for $j \geq M_t$} 
The signal $X_{j}^{(1)}$ is transmitted by the $M_{t}$ transmitters from the transmit set $\sT_{j} = j\uparrow M_{t}$. The corresponding beam $\bv^{(1)}_{j}$ is designed to avoid the interference at the first $M_{t}-1$ receivers $\sI = 1 \uparrow (M_{t}-1)$. Therefore, we see that the contribution of $X_{j}^{(1)}$ at receiver $i$ is given by 
\be{eq:zf-km-1} 
\begin{split} 
g_{ij}^{(1)} = \det{\bH(\sA,\sB)} 
\end{split} 
\ee 
where 
\[
\begin{split} 
\sA & = \{1,2,\cdots,M_{t}-1,i\} \\ 
\sB & = \{j,j+1,\cdots,j+M_{t}-1\}. 
\end{split} 
\]
\subsubsection{Design of transmit beams $\bv_{j}^{(1)}$ and $\bv_{j}^{(2)}$ for $j < M_t$}
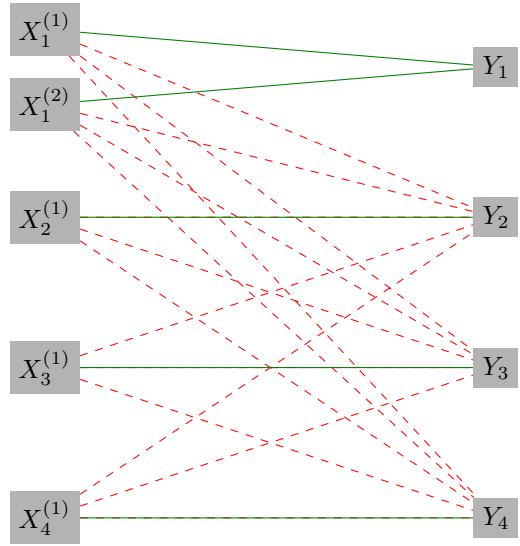
\begin{figure}[th!] 
\centering 
\begin{tikzpicture} 
        \tikzstyle{tx}=[rectangle,fill=black!30] 
        \tikzstyle{rx}=[rectangle,fill=black!30] 
        \def\K{4}\def\M{2}\def\Mm{1} \def\sep{6} 
        \colorlet{dircolor}{green!50!black} 
        \colorlet{intcolor}{red} 
 
        \foreach \name / \i in {1,...,\Mm}{ 
               \node[tx] (t1-\name) at (0,-2*\i) {$X^{(1)}_\i$}; 
                 \node[tx] (t2-\name) at (0,-1-2*\i) {$X^{(2)}_\i$}; 
                \node[rx](r-\name) at (\sep,-0.5-2*\i) {$Y_\i$};} 
        \foreach \name / \i in {\M,...,\K}{ 
               \node[tx] (t-\name) at (0,-0.5-2*\i) {$X^{(1)}_\i$}; 
                 \node[rx] (r-\name) at (\sep,-0.5-2*\i) {$Y_\i$};} 
        \foreach \j in {1,...,\Mm} 
                \foreach \i in {\M,...,\K}{ 
                           \path[dashed,intcolor] (t1-\j) edge (r-\i); 
                        \path[dashed,intcolor] (t2-\j) edge (r-\i);} 
        \foreach \j in {\M,...,\K} 
                \foreach \i in {\M,...,\K} 
                           \path[dashed,intcolor] (t-\j) edge (r-\i); 
        \foreach \name / \i in {1,...,\Mm}{ 
               \path[dircolor] (t1-\i) edge (r-\i); 
                 \path[dircolor] (t2-\i) edge (r-\i);} 
        \foreach \name / \i in {\M,...,\K}{ 
               \path[dircolor] (t-\i) edge (r-\i); 
                } 
\end{tikzpicture} 
\caption{The derived channel in Section~\ref{sec:comp-tx} when $K = 4$ and $M_t = 2$. The thick green lines indicate the links carrying signal. The dashed red lines indicate the links carrying interference.} 
\label{fig:dc-km} 
\end{figure}

The signals $X_{j}^{(1)}$ and $X_{j}^{(2)}$ are transmitted by the $M_{t}$ transmitters from the transmit set $\sT_{j} = 1\uparrow M_{t}$. They must avoid interference at the $M_{t}-2$ receivers 
\[
\sI = \{1,2,\cdots,j-1,j+1,\cdots,M_{t}-1\}. 
\]
Since we only need to avoid interference at $M_{t}-2$ receivers, it is sufficient to transmit each signal from $M_{t}-1$ number of transmitters. We use the first $M_{t}-1$ antennas of the transmit set $\sT_{j}$ to transmit $X_{j}^{(1)}$, and the last $M_{t}-1$ antennas of the transmit set $\sT_{j}$ to transmit $X_{j}^{(2)}$. Thus, we obtain 
\be{eq:zf-km-2} 
\begin{split} 
g_{ij}^{(1)} = \det{\bH(\sA,\sB_{1})} \\ 
g_{ij}^{(2)} = \det{\bH(\sA,\sB_{2})} \\ 
\end{split} 
\ee 
where 
\[
\begin{split} 
\sA & = \{1,2,\cdots,j-1,j+1,M_{t}-1,i\} \\ 
\sB_{1} & = \{j,j+1,\cdots,j+M_{t}-2\} \\ 
\sB_{2} & = \{j+1,j+1,\cdots,j+M_{t}-1\}. 
\end{split} 
\]
Thus, the derived channel \eqref{eq:dc-km} can be simplified as 
\be{eq:dc-km2} 
\begin{split}
Y_{i} & = g_{ii}^{(1)} X_{j}^{(1)} + g_{ii}^{(2)}X_{j}^{(2)} + Z_{i}, \, 1 \leq i < M_{t} \\
Y_{i} & = \sum_{j = 1}^{K} g_{ij}^{(1)} X_{j}^{(1)} + \sum_{j=1}^{M_{t}-1} g_{ij}^{(2)}X_{j}^{(2)} + Z_{i}, \, M_{t} \leq i \leq K
\end{split} 
\ee 
where the derived channel coefficients are as described in \eqref{eq:zf-km-1} and \eqref{eq:zf-km-2}. Figure~\ref{fig:dc-km} provides a description of the derived channel for the special case of $K = 4$ and $M_t = 2$. We note that the derived channel in this section is a not a generalization, and does not specialize to the derived channel in Section~\ref{sec:comp-tx-mk2} when $M_{t} = K-2$. In fact, the achievable scheme in this section achieves fewer DoF compared to the optimal $\frac{KM_{t}}{M_{t}+1}$ DoF achieved in Section~\ref{sec:comp-tx-mk2}.
 
\subsection{Asymptotic Interference Alignment}  
In this section, we consider $L$ parallel derived channels, and propose a scheme achieving a DoF arbitrary close to $(K+M_{t}-1)/2$ in the limit $L \rightarrow \infty$. As in Section~\ref{sec:comp-tx-mk2-ia}, we can combine $L$ parallel derived channels \eqref{eq:dc-km2} and express them together as
\[
\begin{split}
\uy_{i} & = \bG_{ii}^{(1)}\ux_{j}^{(1)} + \bG_{ii}^{(2)}\ux_{j}^{(2)} + \uz_{i}, \, 1 \leq i < M_{t} \\
\uy_{i} & = \sum_{j = 1}^{K} \bG_{ij}^{(1)} \ux_{j}^{(1)} + \sum_{j=1}^{M_{t}-1} \bG_{ij}^{(2)}\ux_{j}^{(2)} + \uz_{i}, \, M_{t} \leq i \leq K
\end{split} 
\]
where $\ux^{(m)}_{j}, \uy_{i}$ and $\uz_{i}$ are $L \times 1$ column vectors and $\bG_{ij}^{(m)}$ is $L \times L$ diagonal channel transfer matrix given by
\[
\bG_{ij}^{(m)} = \mat{g_{ij}^{(m)}(1) \\ & g_{ij}^{(m)}(2) \\ & & \ddots \\ & & & g_{ij}^{(m)}(L)}.
\] 
As in Section~\ref{sec:comp-tx-mk2-ia}, we use $\bV$, defined as the order$-n$ CJ subspace generated by the channel matrices carrying interference
\be{eq:comp-tx-mk-int-mats}
\begin{split} 
& \{\bbG_{ij}^{(1)}, \bbG_{ij}^{(2)}: i \geq M_t, j < M_{t}\} \cup \{\bbG_{ij}^{(1)}: i \neq j \geq M_t\}
\end{split} 
\ee
as the transmit beamforming matrix at every transmitter of the derived channel. The first $M_{t}-1$ receivers do not see any interference. Therefore, for each $k < M_t$, the receiver $k$ can decode all the desired streams free of interference if the matrix 
\[
\begin{split} 
\bM_{k} = \mat{\bG_{kk}^{(1)}\bV & \bG_{kk}^{(2)}\bV} 
\end{split} 
\]
has full column rank. Assuming that the number of rows in $\bM_{k}$, equal to the number of parallel channels $L$, is greater than or equal to the number of columns, i.e., $L \geq 2|\bV|$, the matrix $\bM_k$ has full column rank for generic (original) channel coefficients $\{h_{ij}(\ell)\}$ if the following claim is true. See Corollary~\ref{corr:app-ag-fullrank} in Section~\ref{sec:app-ag} for an explanation.  
\begin{claim} \label{claim:km-1} 
For each $k < M_t$, the polynomials denoted by the variables 
\be{eq:gij-1} 
\begin{split} 
\{g_{kk}^{(1)}, g_{kk}^{(2)}\} & \cup \{g_{ij}^{(1)}, g_{ij}^{(2)} : i \geq M_t, j < M_t\} \cup \{g_{ij}^{(1)} : i \neq j \geq M_t\} 
\end{split} 
\ee 
are algebraically independent.  
\end{claim} 
For each $k \geq M_t$, the interference seen at receiver $k$ is limited to the order$-(n+1)$ CJ subspace, denoted by $\int$. Therefore, the receiver $k$ can decode all the desired streams free of interference if the matrix 
\[
\begin{split} 
\bM_{k} & \ = \mat{\bG_{kk}^{(1)}\bV & \int} 
\end{split} 
\]
has full column rank. Assuming that the number of rows is greater than or equal to the number of columns, i.e., $L \geq |\bV| + |\int|$, the matrix $\bM_k$ has full column rank for generic (original) channel coefficients $\{h_{ij}(t)\}$ if the following claim is true.  
\begin{claim} \label{claim:km-2} 
For each $k \geq M_t$, the polynomials denoted by the variables
\be{eq:gii-2} 
\begin{split} 
\{g_{kk}^{(1)}\} & \cup \{g_{ij}^{(1)}, g_{ij}^{(2)} : i \geq M_t, j < M_t\} \cup \{g_{ij}^{(1)} : i \geq M_t, j \geq M_t\} 
\end{split} 
\ee 
are algebraically independent.  
\end{claim} 
 
To satisfy the requirements on $L$, we choose $L$ as
\[
L = \max(2|\bV|,|\bV| + |\int|) = |\bV| + |\int|.
\] 
Observe that
\[
\begin{split} 
|\bV| = & \ {N + n \choose n} \text{ and } 
|\int| = {N+n+1 \choose n+1}
\end{split} 
\]
where $N$ is the number of matrices \eqref{eq:comp-tx-mk-int-mats} used to generate the CJ subspace, and is given by
\be{eq:comp-tx-mk-N}
\begin{split}
N & = 2(K-M_{t}+1)(M_{t}-1) + (K-M_{t}+1)(K-M_{t}) \\
& = (K-M_{t}+1)(K + M_{t} - 2).
\end{split}
\ee
Therefore, the achievable DoF is given by 
\[
\begin{split} 
\dof(K,M_{t},1,L) & \geq \frac{2(M_{t}-1)|\bV| + (K-M_{t}+1)|\bV|}{L} \\ 
& = \frac{(K+M_{t}-1)|\bV|}{|\bV| + |\int|} \\ 
& = \frac{K+M_{t}-1}{2 + \frac{N}{n+1}}.
\end{split} 
\] 
Therefore, we obtain that
\[
\begin{split}
\dof(K,M_t,1) & = \limsup_{L \rightarrow \infty}\dof(K,M_t,1,L) \\
&  \geq \lim_{n \rightarrow \infty} \frac{K+M_{t}-1}{2 + \frac{N}{n+1}} \\
& = \frac{K + M_{t} - 1}{2}.
\end{split}
\]
\subsection{Proof of Algebraic Independence} \label{sec:comp-tx-mk-ai}
As in Section~\ref{sec:comp-tx-mk2-ag}, we use the Jacobian criterion to prove Claims~\ref{claim:km-1} and \ref{claim:km-2}.  Recall that each derived channel coefficient is a polynomial in $K^{2}$ variables $\{h_{ij}: 1 \leq i,j, \leq K\}$. Let $\bg$ denote the vector consisting of the polynomials specified by the derived channel coefficients in the respective claims. The exact description of the polynomials can be obtained from \eqref{eq:zf-km-1} and \eqref{eq:zf-km-2} in Section~\ref{sec:zf-km}. The number of polynomials in Claims~\ref{claim:km-1} and \ref{claim:km-2} is equal to $N+2$ and $N+1$, respectively, where $N$ is given by \eqref{eq:comp-tx-mk-N}. From Lemma~\ref{lemma:app-ag-jacob} in Section~\ref{sec:app-ag}, we see that a collection of polynomials is algebraically independent if and only if the corresponding Jacobian matrix has full row rank. It can be easily verified that $N+2 \leq K^{2}$, and hence $N+1 \leq K^{2}$, for any $K$ and $M_{t}$, which is a necessary condition for the corresponding Jacobian matrices to have full row rank. It is easy to verify that the Jacobian matrices corresponding to the polynomials in Claims~\ref{claim:km-1} and \ref{claim:km-2} have full row rank using symbolic toolbox of MATLAB for any fixed $K$ and $M_{t}$. In particular, we verified that the Jacobian matrices have full row rank for all values of $M_{t} < K \leq 9$. 
\section{Conclusion} \label{sec:concl}
We studied the problem of characterizing the DoF of the $K$-user CoMP channel with a transmit cooperation order of $M_t$ and a receive cooperation order of $M_r$.  Theorem~\ref{thm:comp-fulldof} says that the DoF equals its maximal value $K$ if and only if $M_t + M_r \geq K+1$. It was known from previous work that the maximum $K$ DoF is achievable by perfect cooperation at either the transmitters or the receivers, i.e., $M_t = K$ or $M_r = K$. Theorem~\ref{thm:comp-fulldof} says that it is possible to achieve the maximum $K$ DoF with only partial cooperation at both the transmitters and receivers. Theorem~\ref{thm:comp-tx} says that the DoF with only CoMP transmission is roughly equal to $\frac{K + M_t - 1}{2}$. We could 
verify using MATLAB that the Theorem holds true for all values of $M_{t} < K < 10$, but we conjecture that the Theorem holds true for all $K$ and $M_{t}$.

The outer bound in Theorem~\ref{thm:comp-ob-sumdof} states that the DoF is bounded above by $\lt\lceil\frac{K+M_t+M_r-2}{2}\rt\rceil$. Since the interference channel with no cooperation has $\frac{K}{2}$ DoF, this outer bound implies that CoMP transmission and reception does not yield significant DoF improvements in the large user regime where $K$ is large compared to $M_t$ and $M_r$. It is not clear if this pessimistic insight is fundamental or is an artifact of the choice of transmit sets \eqref{eq:tx-sets} and receive sets \eqref{eq:rx-sets}. The outer bound in Theorem~\ref{thm:comp-ob-sumdof} fails if we allow the transmit and receive sets to be arbitrary but satisfying the cooperation order constraints, i.e, $|\sT_k| \leq M_t$ and $|\sR_k| \leq M_r$. For the special case of CoMP transmission, i.e., with $M_r =1$, we can use Theorem~\ref{thm:comp-dof-ob-region} to show that the DoF is outer bounded by $\frac{KM_t}{M_t+1}$ no matter how the transmit sets are chosen. Theorem~\ref{thm:comp-tx-mk2} says that this DoF is achieved using spiral transmit sets \eqref{eq:tx-sets} when $M_{t} = K-2$. In general, this may not be true, and the problem of determining the DoF with arbitrary transmit sets remain open. 

The achievability proofs in this paper are heavily dependent on concepts from algebraic geometry, specifically on the notion of algebraic independence of rational functions. Similar tools have recently been used in \cite{yetis2010feasibility,bresler2011settling,razaviyayn2011degrees} to determine the feasibility of interference alignment in MIMO interference channels with no cooperation. We believe that these tools further our understanding of the DoF of wireless channels, and have the potential to settle many other feasibility questions.



\appendix
\subsection{Proof of Lemma~\ref{lemma:fullrank}} \label{sec:app-fullrank}
We have already proved that $\bM$ does not have full column rank if there exists an annihilating polynomial $F$ of the form \eqref{eq:F-as}. We now prove the converse; i.e., we assume that there does not exist an annihilating polynomial of the form \eqref{eq:F-as}, and prove that the matrix $\bM$ has full column rank for generic realizations of the variables \eqref{eq:t-q}. Without any loss of generality, we assume that $p = q$. Otherwise, we can work with the $q \times q$ submatrix obtained after deleting the last $q - p$ rows.  
 
Consider expanding the determinant $\det{\bM}$ in terms of the variables \eqref{eq:t-q}. Since the variables $\bs(1), \bs(2), \cdots, \bs(q)$ are rational functions of $\bt(1), \bt(2), \cdots, \bt(q)$ respectively, the determinant is also a rational function; i.e., 
\[ 
\det{\bM} = \frac{d_{1}(\bt(1),\bt(2),\cdots,\bt(q))}{d_{2}(\bt(1),\bt(2),\cdots,\bt(q))}. 
\]
The determinant can either be identically equal to zero, or a nonzero function. If the determinant is a nonzero function, then $\bM$ has full column rank for generic realizations of the variables \eqref{eq:t-q} because $\bM$ is rank deficient only when $d_{1}(\bt(1),\bt 2),\cdots,\bt(q)) = 0$ or when $(\bt(1),\bt(2),\cdots,\bt(q))$ belongs to the affine variety $V(d_{1}) \subsetneq \mathbb{C}^{nq}$ generated by the polynomial $d_{1}$.  
 
Therefore, it remains to prove that $\det{M}$ is not identically equal to zero under the assumption that no annihilating polynomial $F$ of the form \eqref{eq:F-as} exists. We prove this claim by induction on $q$. The claim is trivial to check for $q = 1$. We now prove the induction step. We may assume that the determinant of the $(q-1)\times(q-1)$ submatrix $\tilde{\bM}$, obtained after deleting the last row and column, is a nonzero function in $(\bt(1),\bt(2),\cdots,\bt(q-1))$. Therefore, there must exist specific realizations  
\be{eq:q-1} 
(\bt(1),\bt(2),\cdots,\bt(q-1)) = (\ba(1),\ba(2),\cdots,\ba(q-1))
\ee 
such that $\tilde{\bM}$ has full rank. Consider the matrix $\bM^{*}(\bt)$ obtained from $\tilde{\bM}$ by setting $\bt(q) = \bt$ for each $\bt \in \mathbb{C}^{n}$. If $\det{\bM}$ is identically equal to zero, then the matrix $\bM^{*}(\bt)$ must be rank deficient for all $\bt$; i.e., there must exist $\bc(\bt) \neq \bo$ such that $\bM^{*}(\bt)\bc(\bt) = \bo$ for each $\bt \in \mathbb{C}^{n}$. Since the first $q-1$ rows are linearly independent and do not depend on $\bt$, the vector $\bc(\bt) = \bc^{*}$ is unique (up to a scaling factor) and is determined by \eqref{eq:q-1}. Therefore, we have that $\bM^{*}(\bt)\bc^{*} = \bo$ for each $\bt \in \mathbb{C}^{n}$. By expanding the last row of $\bM^{*}(\bt)\bc^{*} = \bo$, we obtain
\[
\sum_{i=1}^{q}c^{*}_{i} \mathbf{f}(\bt)^{\ba_{i}} = 0. 
\]
This is a contradiction since we assumed that no annihilating polynomial of the form \eqref{eq:F-as} exists. Therefore, $\det{\bM}$ is not identically equal to zero and hence $\bM$ has full rank for generic realizations of the variables \eqref{eq:t-q}.
\subsection{Appendix: Proof of Claim~\ref{claim:comp-jacob}} \label{sec:comp-claim-jacob}

In this section, we complete the proof of Theorem~\ref{thm:comp-tx-mk2} by show that the determinant of the submatrix \eqref{eq:subJ} evaluated at the point $\bH = \bA$ is equal to $\pm 1$. Recall that
\[
\begin{split} 
& \bg_0 = \lt(g_{11}^{(1)},g_{11}^{(2)}, \cdots, g_{11}^{(M_t)} \rt) = \bH(1,\sT_{1})\bH(\sT_{2},\sT_{1})^{-1}\\  
& \bg_i = \lt(g_{i,i+1}^{(1)},g_{i,i+1}^{(2)}, \cdots, g_{i,i+1}^{(M_t)} \rt) = \bH(i,\sT_{i+1})\bH(\sT_{i+2},\sT_{i+1})^{-1} \\ 
& \bh_0 = \lt(h_{11}, h_{22}, \cdots, h_{M_tM_t}\rt) \\  
& \bh_i = \lt(h_{i,i+1}, h_{i,i+2}, \cdots, h_{i,K},h_{i,1}, \cdots,h_{i,i-2}\rt) = \bH(i,\sT_{i+1})
\end{split} 
\]
where the transmit set $\sT_{i}$ is given by
\[
\sT_{i} = i \uparrow (K-2) = \{i,i+1,\cdots,i+K-2\}.
\]
Let $\bJ[\bg_i;\bh_j]$ denote the submatrix of the Jacobian matrix with rows corresponding to the variables $\bg_i$ and columns corresponding to the variables $\bh_j$. Then, the submatrix \eqref{eq:subJ} can be expressed as 
\be{eq:subJ2} 
\mat{\bJ[\bg_0;\bh_0] & \cdots & \bJ[\bg_0;\bh_K] \\ 
\vdots & \ddots & \vdots \\ 
\bJ[\bg_K;\bh_0] & \cdots & \bJ[\bg_K;\bh_K] 
}. 
\ee 
Differentiating $\bg_i  = \bH(i,\sT_{i+1})\bH(\sT_{i+2},\sT_{i+1})^{-1}$ at $\bH = \bA$, we get  
\be{eq:dgi} 
\begin{split} 
& d\bg_i = d\bH(i,\sT_{i+1})\bA(\sT_{i+2},\sT_{i+1})^{-1} \\ 
& ~~~~~~~~ - \bA(i,\sT_{i+1})\bA(\sT_{i+2},\sT_{i+1})^{-1}d\bH(\sT_{i+2},\sT_{i+1})\bA(\sT_{i+2},\sT_{i+1})^{-1} .
\end{split} 
\ee 
The matrix $\bA$ is chosen to satisfy 
\[
\begin{split} 
\bA(i,\sT_{i+1}) = & \ \bo \\ 
\bA(\sT_{i+2},\sT_{i+1}) = & \ \bB, 
\end{split} 
\]
where $\bB$ is the $M_t \times M_t$ matrix with all the diagonal and the superdiagonal entries being equal to $1$. Note that $\det{\bB}=1$. For the special case of $K = 4$ and $M_t = 2$, the matrix $\bB$ is given by 
\[
\bB = \mat{1 & 1 \\ 0 & 1}. 
\]
Therefore, \eqref{eq:dgi} can be simplified as 
\[
\begin{split} 
d\bg_i = d\bH(i,\sT_{i+1})\bB^{-1} = d\bh_i\bB^{-1}.
\end{split} 
\]
Equivalently, for each $i \geq 1$, we have 
\[
\begin{split} 
\bJ[\bg_i;\bh_i] = & \ \bB^{-\top} \\ 
\bJ[\bg_i;\bh_j] = & \ \bo, \forall j \neq i. 
\end{split} 
\]
Hence, the determinant of the submatrix \eqref{eq:subJ2} is equal to 
\[
\det{\bJ[\bg_0;\bh_0]}/ (\det{\bB})^{K} = \det{\bJ[\bg_0;\bh_0]}. 
\]
We now show that $\det{\bJ[\bg_0;\bh_0]} = \pm1$. Recall from Section~\ref{sec:comp-tx-zf} that $\bg_0$ is related to $\bH$ as 
\[
\bg_0 = \lt(g_{11}^{(1)},g_{11}^{(2)}, \cdots, g_{11}^{(M_t)} \rt) = \bH(1,\sT_{1})\bH(\sT_{2},\sT_{1})^{-1}. 
\]
Differentiating $\bg_0 = \bH(1,\sT_{1})\bH(\sT_{2},\sT_{1})^{-1}$ at $\bH = \bA$, we get  
\[
\begin{split} 
d\bg_0 & = d\bH(1,\sT_{1})\bA(\sT_{2},\sT_{1})^{-1} \\ 
& ~~~~~~~~~  - \bA(1,\sT_{1})\bA(\sT_{2},\sT_{1})^{-1}d\bH(\sT_{2},\sT_{1})\bA(\sT_{2},\sT_{1})^{-1} \\ 
& = d\bH(1,\sT_{1})\bB^{-1} - \bA(1,\sT_{1})\bB^{-1}d\bH(\sT_{2},\sT_{1})\bB^{-1}.
\end{split} 
\]
Now, observe that 
\[
\begin{split} 
 \bA(1,\sT_{1})\bB^{-1} = & \ \mat{1 & 0 & \cdots & 0}\bB^{-1} \\ 
 = & \ \mat{1 & -1 & 1 & -1 & \cdots}. 
\end{split} 
\]
Therefore, we get 
\[ 
\begin{split} 
d\bg_0 \bB = & \ d\bH(1,\sT_{1}) - \bA(1,\sT_{1})\bB^{-1}d\bH(\sT_{2},\sT_{1}) \\ 
= & \mat{dh_{11} & dh_{12} & \cdots & dh_{1,K-2}} \\ 
& - \mat{dh_{21} & dh_{22} & \cdots & dh_{2,K-2}} \\ 
&  + \mat{dh_{31} & dh_{32} & \cdots & dh_{3,K-2}} \\ 
& ~ \vdots \\ 
&  (-1)^{K-1} \mat{dh_{K-2,1} & dh_{K-2,2} & \cdots & dh_{K-2,K-2}} \\ 
&  (-1)^{K}  \mat{dh_{K-1,1} & dh_{K-1,2} & \cdots & dh_{K-1,K-2}}. 
\end{split} 
\] 
To determine $\bJ[\bg_0;\bh_0]$, we are only interested in the partial derivatives with respect to the variables $h_{11}, h_{22}, \cdots, h_{K-2,K-2}$. The contribution of  $d\bh_0$ in $d\bg_0$ is given by 
\[
\begin{split} 
& \ \mat{dh_{11} & - dh_{22} & dh_{33} & - dh_{44} & \cdots}\bB^{-1} \\ 
& ~~~~ = d\bh_0\mat{1 \\ & -1 \\ & & 1 \\ &&&\ddots}\bB^{-1} 
\end{split} 
\]
which implies that 
\[
\bJ[\bg_0;\bh_0] = \bB^{-\top}\mat{1 \\ & -1 \\ & & 1 \\ &&&\ddots}. 
\]
Hence, $\det{\bJ[\bg_0;\bh_0]} = \pm \det{\bB} = \pm1$. 

\bibliographystyle{IEEEtran} 
\bibliography{thesisrefs} 
\end{document}

%% file: notation.tex
\newtheorem{lemma}{Lemma}
\newtheorem{ques}{Question}
\newtheorem{claim}{Claim}
\newtheorem{theorem}{Theorem}
\newtheorem{corollary}{Corollary}
\newtheorem{definition}{Definition}
\newtheorem{remark}{Remark}
\newtheorem{conjecture}{Conjecture}
\newtheorem{example}{Example}
\newcommand{\mat}[2][cccccccccccccccccccccccccccccccccc]{\left[\begin{array}{#1}#2\\\end{array}\right]}
\newcommand{\be}[1]{\begin{equation} \label{#1}}
\newcommand{\ee}{\end{equation}}
\newcommand{\lt}{\left}
\newcommand{\rt}{\right}
\newcommand{\entropy}[1]{\mathsf{h}\left(#1\right)}
\newcommand{\Entropy}[1]{\mathsf{H}\left(#1\right)}
\newcommand{\ent}[1]{\mathsf{h}\left(#1\right)}
\newcommand{\Ent}[1]{\mathsf{H}\left(#1\right)}
\newcommand{\mi}[1]{\mathsf{I}\left(#1\right)}
\newcommand{\tr}[1]{\mathsf{Tr}\left(#1\right)}
\renewcommand{\int}{\mathbf{INT}}
\newcommand{\mb}[1]{\mathbf{#1}}
\newcommand{\bo}{\mathbf{0}}
\newcommand{\ba}{\mathbf{a}}
\newcommand{\bc}{\mathbf{c}}
\newcommand{\bg}{\mathbf{g}}
\newcommand{\bh}{\mathbf{h}}
\newcommand{\bp}{\mathbf{p}}
\newcommand{\bq}{\mathbf{q}}
\newcommand{\br}{\mathbf{r}}
\newcommand{\bs}{\mathbf{s}}
\newcommand{\bt}{\mathbf{t}}
\newcommand{\bu}{\mathbf{u}}
\newcommand{\bv}{\mathbf{v}}
\newcommand{\bw}{\mathbf{w}}
\newcommand{\bx}{\mathbf{x}}
\newcommand{\by}{\mathbf{y}}
\newcommand{\bz}{\mathbf{z}}

\newcommand{\bA}{\mathbf{A}}
\newcommand{\bB}{\mathbf{B}}
\newcommand{\bG}{\mathbf{G}}
\newcommand{\bH}{\mathbf{H}}
\newcommand{\bI}{\mathbf{I}}
\newcommand{\bJ}{\mathbf{J}}
\newcommand{\bM}{\mathbf{M}}
\newcommand{\bR}{\mathbf{R}}
\newcommand{\bS}{\mathbf{S}}
\newcommand{\bU}{\mathbf{U}}
\newcommand{\bV}{\mathbf{V}}
\newcommand{\tbU}{\tilde{\mathbf{U}}}

\newcommand{\bbs}{\bar{\mathbf{s}}}
\newcommand{\bbx}{\bar{\mathbf{x}}}
\newcommand{\bby}{\bar{\mathbf{y}}}
\newcommand{\bbz}{\bar{\mathbf{z}}}
\newcommand{\bbG}{\bar{\mathbf{G}}}
\newcommand{\bbV}{\bar{\mathbf{V}}}
\newcommand{\bbI}{\bar{\mathbf{I}}}

\newcommand{\uy}{\underline{Y}}
\newcommand{\ux}{\underline{X}}
\newcommand{\uz}{\underline{Z}}
\newcommand{\us}{\underline{S}}
\newcommand{\ue}{\underline{E}}
\newcommand{\un}{\underline{N}}

\newcommand{\ubx}{\underline{\bar{X}}}
\newcommand{\uby}{\underline{\bar{Y}}}
\newcommand{\ubz}{\underline{\bar{Z}}}
\newcommand{\sA}{\mathcal{A}}
\newcommand{\sB}{\mathcal{B}}
\newcommand{\sC}{\mathcal{C}}
\newcommand{\sD}{\mathcal{D}}
\newcommand{\sE}{\mathcal{E}}
\newcommand{\sF}{\mathcal{F}}
\newcommand{\sG}{\mathcal{G}}
\newcommand{\sH}{\mathcal{H}}
\newcommand{\sI}{\mathcal{I}}
\newcommand{\sM}{\mathcal{M}}
\newcommand{\sK}{\mathcal{K}}
\newcommand{\sS}{\mathcal{S}}
\newcommand{\sT}{\mathcal{T}}
\newcommand{\sR}{\mathcal{R}}
\newcommand{\sW}{\mathcal{W}}
\newcommand{\co}[1]{\mathtt{\sim}\lt\{#1\rt\}}
\newcommand{\spm}[2]{\mathbb{SM}_{#1,#2}}
\newcommand{\cF}{\mathcal{F}}
\newcommand{\cG}{\mathcal{G}}
\newcommand{\cH}{\mathcal{H}}
\newcommand{\ch}{\mathcal{h}}
\newcommand{\dof}{\text{DoF}}
\newcommand{\pudof}{\text{P.U.DoF}}